\documentclass[runningheads]{llncs}
\usepackage{graphicx}
\usepackage{times}

\usepackage{soul}
\usepackage{url}
\usepackage[hidelinks,breaklinks]{hyperref}
\usepackage[utf8]{inputenc}
\usepackage[small]{caption}
\usepackage{graphicx}
\usepackage{amsmath}
\usepackage{amssymb}
\usepackage{booktabs}
\usepackage{subfigure}
\usepackage{color}
\usepackage{multirow}
\usepackage{threeparttable}
\spnewtheorem*{myproblem}{Problem}{\bf}{\itshape}

\newcommand{\hhtne}{$\text{H}^2\text{TNE}$}
\begin{document}
\title{\hhtne~: Temporal Heterogeneous Information Network Embedding in Hyperbolic Spaces}
%
% \titlerunning{\hhtne~: hyperbolic heterogeneous temporal network embedding}
\titlerunning{Temporal HIN Embedding in Hyperbolic Spaces}
% If the paper title is too long for the running head, you can set
% an abbreviated paper title here
%

\author{Qijie Bai\inst{1,3}\and
Jiawen Guo\inst{2,3}\and
Haiwei Zhang\inst{1,2,3}\thanks{Corresponding author.}\and
Changli Nie\inst{1,3}\and\\
Lin Zhang\inst{1,3}\and
Xiaojie Yuan\inst{1,2,3}}
\authorrunning{Q. Bai et al.}
% First names are abbreviated in the running head.
% If there are more than two authors, 'et al.' is used.
%
\institute{College of Computer Science, Nankai University, Tianjin, China\and
	College of Cyber Science, Nankai University, Tianjin, China\and
	TJ Key Laboratory of NDST, Nankai University, Tianjin, China\\
	\email{
    	\{\href{mailto:qijie.bai@mail.nankai.edu.cn}{qijie.bai},\href{mailto:nie_cl@mail.nankai.edu.cn}{nie\_cl}\}@mail.nankai.edu.cn,
    	\{\href{mailto:guojiawen@dbis.nankai.edu.cn}{guojiawen},\href{mailto:zhanglin@dbis.nankai.edu.cn}{zhanglin}\}@dbis.nankai.edu.cn,
    % 	\href{mailto:linzhang@stumail.neu.edu.cn}{linzhang@stumail.neu.edu.cn},
    	\{\href{mailto:zhhaiwei@nankai.edu.cn}{zhhaiwei},\href{mailto:yuanxj@nankai.edu.cn}{yuanxj}\}@nankai.edu.cn
	}
}

\maketitle              % typeset the header of the contribution
\begin{abstract}
    \textit{Temporal heterogeneous information network} (temporal HIN) embedding, aiming to represent various types of nodes of different timestamps into low-dimensional spaces while preserving structural and semantic information, is of vital importance in diverse real-life tasks. Researchers have made great efforts on temporal HIN embedding in Euclidean spaces and got some considerable achievements. However, there is always a fundamental conflict that many real-world networks show hierarchical property and power-law distribution, and are not isometric of Euclidean spaces. Recently,  representation learning in hyperbolic spaces has been proved to be valid for data with hierarchical and power-law structure. Inspired by this character, we propose a \textit{hyperbolic heterogeneous temporal network embedding} (\hhtne) model for temporal HINs. Specifically, we leverage a temporally and heterogeneously double-constrained random walk strategy to capture the structural and semantic information, and then calculate the embedding by exploiting hyperbolic distance in proximity measurement. Experimental results show that our method has superior performance on temporal link prediction and node classification compared with SOTA models.
    
\end{abstract}

\section{Introduction} \label{sec::intro}
% Recently graph representation learning (also called network embedding) has received great attention on account of its critical role and outstanding performance in graph data analysis~\cite{chenGraphRepresentationLearning2020}. It maps nodes to a low-dimension space while preserving the structural and semantic information of the graph simultaneously, and significantly reduces the time and space complexity in advanced analytic tasks, such as pattern discovery, node classification and link prediction~\cite{cuiSurveyNetworkEmbedding2019}.
 
Heterogeneous information networks (HINs), which are seen as general and simplified knowledge graphs (KGs), are of a ubiquitous structure in various domains. HIN embedding has received great attention in recent years because of its powerful representation abilities for both structural and semantic information in real-world networks~\cite{chenGraphRepresentationLearning2020}. Different types of nodes are mapped into a low-dimension space for diverse downstream analytic tasks, such as pattern matching, node classification and link prediction, within lower time and space complexity~\cite{cuiSurveyNetworkEmbedding2019}.

% \zhhw{\textcolor{blue}{Write something about heterogeneous networks, and then list some existing works for both static and dynamic HIN embedding in Euclidean space.}}

% \zhhw{\textcolor{blue}{Introduce something of hyperbolic embedding as your have written in this part. And then enumerate homo and hetero network embedding methods within hyperbolic space. Finally, summarize something about existing works and emphasize your motivation of this paper.}}

% Most existing works focus on static networks, including static homogeneous networks and static heterogeneous networks. These methods are designed to preserve the topological structure and contextual semantics without considering any temporal information. Deepwalk~\cite{perozziDeepWalkOnlineLearning2014} and node2vec~\cite{groverNode2vecScalableFeature2016} convert irregular networks into node sequences and treat them as corpus in natural language processing. For \textit{heterogeneous information networks} (HINs), a lot methods are also proposed to capture the semantics involved in multiple types of nodes, such as PTE~\cite{tangPTEPredictiveText2015}, HAN~\cite{wangHeterogeneousGraphAttention2019} and MAGNN~\cite{fuMAGNNMetapathAggregated2020}.

Most existing HIN embedding methods focus on static networks. These methods are designed to preserve the topological structure and contextual semantics without considering any temporal information. For example, PTE~\cite{tangPTEPredictiveText2015} embeds text data through heterogeneous text networks. HAN~\cite{wangHeterogeneousGraphAttention2019} and MAGNN~\cite{fuMAGNNMetapathAggregated2020} are proposed to capture the semantics involved in multiple types of nodes.

Complex real-world networks, however, are constantly evolving over time. As a result, how to capture temporal information in dynamic networks becomes more challenging. Many methods are proposed for temporal homogeneous networks like CTDNE~\cite{leeDynamicNodeEmbeddings2020}, while studies on temporal HIN embedding are much less. Most of existing methods for temporal HIN embedding like DHNE~\cite{yinDHNENetworkRepresentation2019}, are non-incremental learning. They are designed to deal with snapshots rather than dynamic networks which keep changing by time. THINE~\cite{huangTemporalHeterogeneousInformation2021} uses attention mechanism and metapath to capture heterogeneous information, and furthermore, Hawkes process is leveraged to simulate the evolution of temporal networks.

\begin{figure}[t]
    \centering
    \subfigure[\scriptsize{Tokyo}]{
        \begin{minipage}[t]{0.47\textwidth}
            \centering
            \includegraphics[width=1\textwidth]{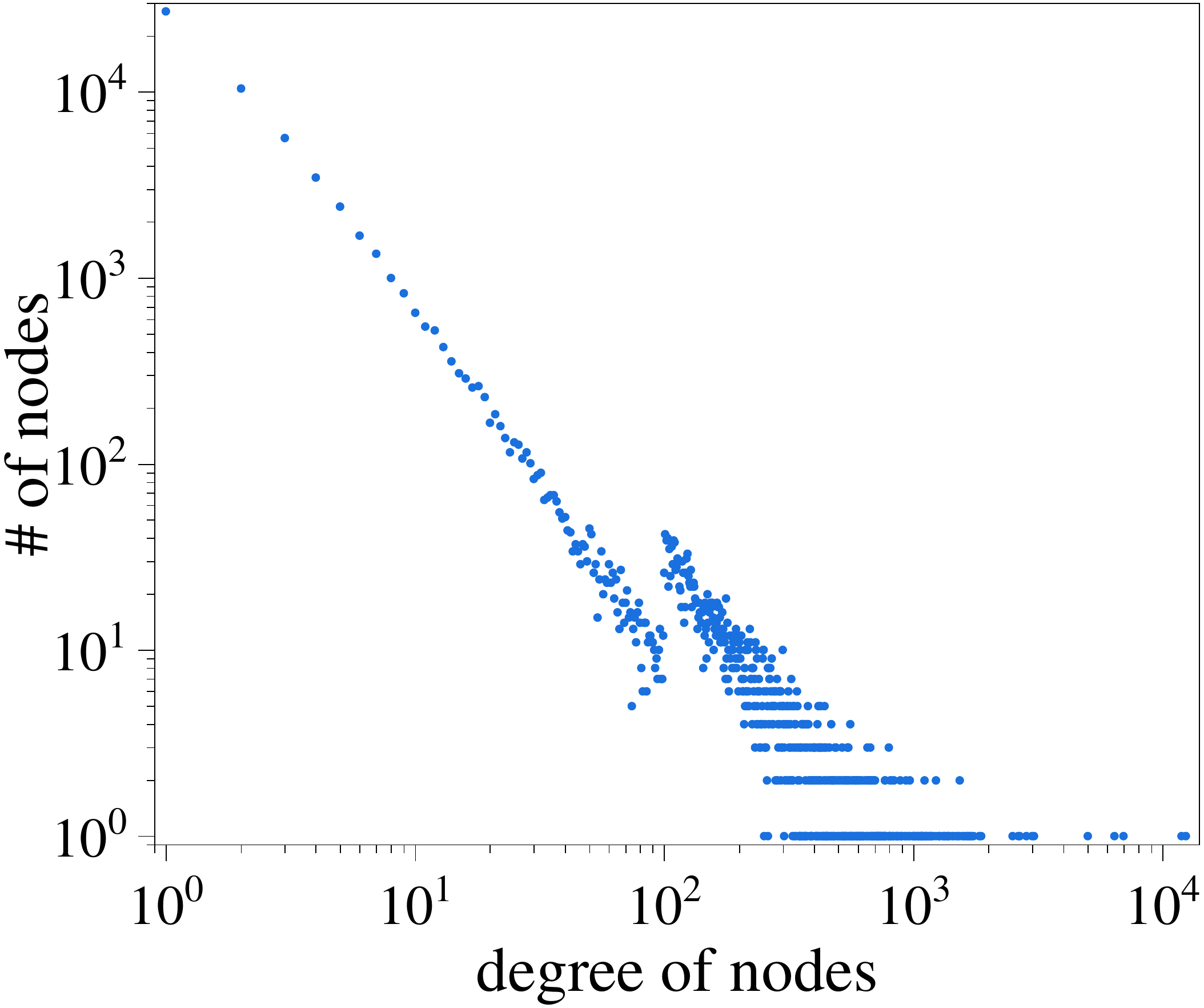}
        \end{minipage}
    }
    \subfigure[\scriptsize{DBLP}]{
        \begin{minipage}[t]{0.47\textwidth}
            \centering
            \includegraphics[width=1\textwidth]{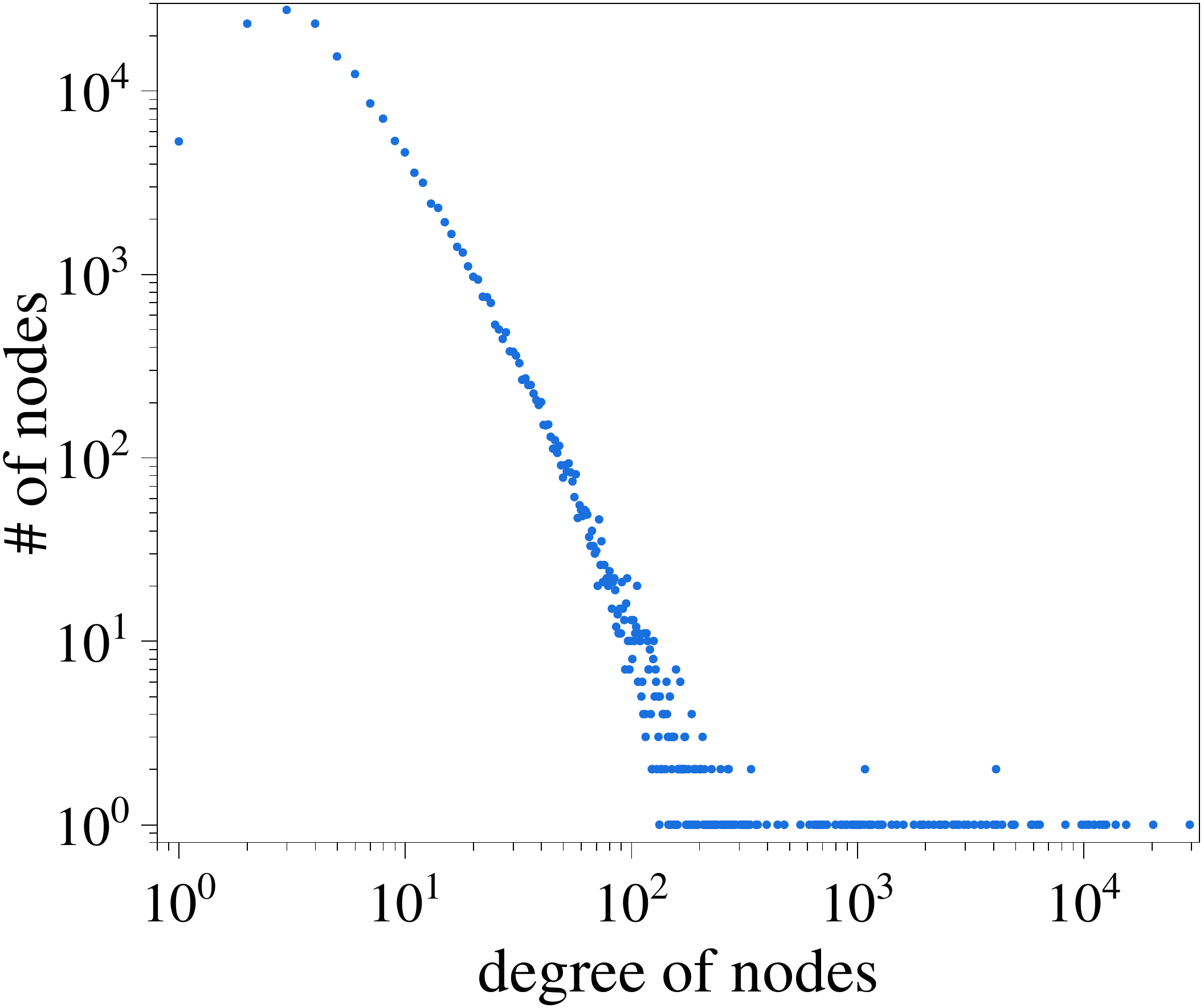}
        \end{minipage}
    }
    % \subfigure[\scriptsize{MovieLens}]{
    %     \begin{minipage}[t]{0.31\textwidth}
    %         \centering
    %         \includegraphics[width=1\textwidth]{pic/movielens_distribution.pdf}
    %     \end{minipage}
    % }
    \caption{The degree distributions of two real-world networks Tokyo and DBLP. The coordinate axes are logarithmic.}
    \label{fig::intro:distribution}
\end{figure}

Although these embedding models have shown great performance in many areas, they are all built upon Euclidean spaces and their representation capacity is inherently limited by the dimension of embedding space. Recently, it has been noticed that complex real-world networks like social networks and many KGs always exhibit non-Euclidean structures~\cite{bronsteinGeometricDeepLearning2017}. As Figure~\ref{fig::intro:distribution} shows, we find that the degrees of nodes follow power-law distributions in most real-world networks, in other words, indicate hyperbolic structures. Hyperbolic spaces are those of constant negative curvature and the areas of disks in hyperbolic spaces grow exponentially with their radius rather than polynomially in Euclidean spaces. Due to the exponential expansion property~\cite{krioukovHyperbolicGeometryComplex2010}, the representation capacity and generalization ability of hyperbolic spaces for data with hierarchical structures or power-law distributions are potentially excellent~\cite{pengHyperbolicDeepNeural2021}. In the past few years, researchers have made some progress in this domain~\cite{nickelPoincareEmbeddingsLearning,wangHyperbolicHeterogeneousInformation2019,zhangHyperbolicGraphAttention2021}. Nevertheless, we find that none of them is designed for temporal HINs.

To this end, we propose \hhtne, a novel hyperbolic heterogeneous temporal network embedding model.  First, we leverage a temporally and heterogeneously double-constrained random walk strategy to capture the topological structure and contextual semantics over time. Then defined in hyperbolic spaces, our model maximizes the proximity between neighbors and minimizes it between negative samples. Moreover, we derive how the optimization process is calculated. Experiments on several real-world datasets show that our \hhtne ~outperforms SOTA methods in two advanced analytic tasks temporal link prediction and node classification. 

The contributions of this paper are summarized as follows:
\begin{itemize}
\item To our best knowledge, we are the first to study general temporal HIN embedding problem in hyperbolic spaces.
% \item We leverage a temporally and heterogeneously double-constrained random walk strategy to capture the structural and semantic information of temporal HINs.
% \item We notice the power-law distributions of real-world datasets and exploit hyperbolic spaces to take full advantage of this feature.
\item We propose a novel temporal HIN embedding model \hhtne, which leverages a temporally and heterogeneously double-constrained random walk strategy to capture the structural and semantic information, and exploits hyperbolic spaces to take full advantage of the power-law distributions for real-world networks.
\item We conduct extensive experiments and the results show our model has better performance than several SOTA methods in node classification and link prediction tasks.
\end{itemize}

The rest of the paper is organized as follows. In Section~\ref{sec::related}, we make a brief but systematical review for related studies. Section~\ref{sec::preli} introduces necessary preliminaries and Section~\ref{sec::model} presents our proposed \hhtne ~from two key modules random walk sampling and hyperbolic embedding. Experiments are described in Section~\ref{sec::expr}. Finally, we conclude our paper in Section~\ref{sec::conclu}.

\section{Related Work}  \label{sec::related}

In this section, we systematically review the existing network embedding methods from three aspects, including traditional network embedding models, deep network embedding models and hyperbolic network embedding models.

% \subsection{Traditional Network Embedding Models}  \label{sec::related:traditional}
\vspace{0.5em}
\noindent
\textbf{Traditional Network Embedding Models.}~
Network embedding aims to map nodes to a low-dimension space without losing structure and semantics of the network. In early researches, because of the unstructured character of networks, different sequential sampling strategies are proposed to simplify the data processing, e.g. Deepwalk~\cite{perozziDeepWalkOnlineLearning2014} and node2vec\cite{groverNode2vecScalableFeature2016}. LINE~\cite{tangLINELargescaleInformation2015} learns node embeddings from the first-order and second-order neighbors. As the researches deepen, heterogeneous and temporal information are taken into account. Metapath2vec~\cite{dongMetapath2vecScalableRepresentation2017} leverages metapath-based random walks and achieves great performance. DynamicTriad~\cite{zhouDynamicNetworkEmbedding} preserves structural information and evolution pattern on network snapshots by triadic closure process. Inspired of above two models, Change2vec~\cite{bianNetworkEmbeddingChange2019a} handles the difference between two snapshots by metapaths and triadic open/closure process.

% \subsection{Deep Network Embedding Models}  \label{sec::related:deep}
\vspace{0.5em}
\noindent
\textbf{Deep Network Embedding Models.}~
With the development of deep learning, many GNN-based embedding models have emerged in recent years. GCN~\cite{kipfSemiSupervisedClassificationGraph2017} aggregates messages from neighbors to update embeddings, and furthermore, GAT~\cite{velickovicGraphAttentionNetworks2018} introduces attention mechanism for aggregation and is competent on inductive tasks. $\rm{M^2DNE}$~\cite{luTemporalNetworkEmbedding2019} describes the temporal evolution of networks in terms of microscopic and macroscopic dynamics. DySAT~\cite{sankarDySATDeepNeural2020} learns on snapshots by multi-head attention and achieves a great performance on link prediction. SHCF~\cite{liSequenceawareHeterogeneousGraph2021} jointly considers sequential information as well as high-order heterogeneous information. THINE\cite{huangTemporalHeterogeneousInformation2021} simulates the dynamic evolution of heterogeneous networks. LIME~\cite{pengLIMELowCostIncremental2021} incrementally trains on temporal HINs and significantly lowers memory resources and computational time. 

% \subsection{Hyperbolic Network Embedding Models}  \label{sec::related:hyperbolic}
\vspace{0.5em}
\noindent
\textbf{Hyperbolic Network Embedding Models.}~
Representation learning in hyperbolic spaces has been applied to network embedding due to the non-Euclidean structures of real-world networks. \cite{nickelPoincareEmbeddingsLearning} learns hierarchical features of networks in the Poincaré ball, while \cite{nickelLearningContinuousHierarchies} discovers pairwise hierarchical relations in Lorentz model. HHNE~\cite{wangHyperbolicHeterogeneousInformation2019} is constructed in hyperbolic spaces on account of the power-law distribution of HINs. \cite{liuHyperbolicGraphNeurala} proposes hyperbolic graph neural networks on graph classification problem, and HGCN~\cite{chamiHyperbolicGraphConvolutional} uses hyperbolic graph convolution networks for node embedding. HAT~\cite{zhangHyperbolicGraphAttention2021} exploits graph attention networks and devises a parallel strategy to improve the efficiency. HTGN~\cite{yangDiscretetimeTemporalNetwork2021} attempts to embed dynamic graphs into hyperbolic geometry. It adopts hyperbolic GCNs to capture spatial features and a hyperbolic temporal contextual attention module to extract the historical information. Besides, h-MDS~\cite{salaRepresentationTradeoffsHyperbolic} provides a precision-dimension trade-off in hyperbolic embedding. \cite{zhangWhereAreWe2021} tells that hyperbolic models are more suited for sparse datasets and greatly outperform Euclidean models when the latent dimension number is small.

All above network embedding models are either built upon Euclidean spaces, or focused on only part of network features. In order to achieve better performance, we propose a hyperbolic embedding model with taking both temporal and heterogeneous information into account.

\section{Preliminaries}  \label{sec::preli}

In this section, we first define the temporal HINs and the problem of temporal HIN embedding. Then some critical properties of hyperbolic geometry are briefly introduced.

\subsection{Temporal HIN Embedding}

Following \cite{leeDynamicNodeEmbeddings2020,chenTutorialNetworkEmbeddings2018,shiSurveyHeterogeneousInformation2017}, HINs and temporal networks are defined traditionally as:

\begin{definition}[HINs]  \label{def::preli:HINs}
A heterogeneous information network is defined as $\mathcal{G}=(\mathcal{V}, \mathcal{E}, \phi, \\ \varphi)$, in which $\mathcal{V}$ and $\mathcal{E}$ are the sets of nodes and edges. Each node $v \in \mathcal{V}$ and each edge $e \in \mathcal{E}$ are associated with mapping functions $\phi : \mathcal{V} \rightarrow \mathcal{L}_{\mathcal{V}}$ and $\varphi : \mathcal{E} \rightarrow \mathcal{L}_{\mathcal{E}}$. $\mathcal{L}_{\mathcal{V}}$ and $\mathcal{L}_{\mathcal{E}}$ denote the sets of node and edge types respectively and satisfy $|\mathcal{L}_{\mathcal{V}}|+|\mathcal{L}_{\mathcal{E}}|>2$.
\end{definition}

Especially, a KG is a natural HIN since it contains different types of objects (e.g. subjects and objects) and links (e.g. properties) \cite{zhengEntitySetExpansion2017}. As a consequence, a general HIN model can be applied into KGs in most cases even with extra information like timestamps.

\begin{definition}[Temporal Networks]  \label{def::preli:tempnets}
A temporal network is defined as $\mathcal{G}=(\mathcal{V}, \mathcal{E}, \tau)$, where $\mathcal{V}$ indicates the set of nodes and $\mathcal{E}$ indicates the set of edges. Each edge $e \in \mathcal{E}$ is associated with the mapping function $\tau: \mathcal{E} \rightarrow \mathcal{T}$, which maps edges to timestamps.
\end{definition}

It is worth noting that most existing temporal network models are designed for intermittent network snapshots, while our work studies the continuous-time dynamic networks. In other words, temporal networks in this paper can be seen as an edge stream and multiple edges may be established between two nodes at different timestamps. Based on Definition~\ref{def::preli:HINs} and Definition~\ref{def::preli:tempnets}, we formalize the temporal HINs as follows:

\begin{definition}[Temporal HINs]  \label{def::preli:tHINs}
A temporal HIN can be formalized as $\mathcal{G}=(\mathcal{V}, \mathcal{E}, \phi, \varphi,\\ \tau)$, in which $\mathcal{V}$ and $\mathcal{E}$ are the sets of nodes and edges. Mapping functions $\phi : \mathcal{V} \rightarrow \mathcal{L}_{\mathcal{V}}$ and $\varphi : \mathcal{E} \rightarrow \mathcal{L}_{\mathcal{E}}$ map nodes and edges into node types and edge types separately. $|\mathcal{L}_{\mathcal{V}}|+|\mathcal{L}_{\mathcal{E}}|>2$ is satisfied. Another mapping function $\tau: \mathcal{E} \rightarrow \mathcal{T}$ maps edges to timestamps. 
\end{definition}

In this paper, we aim to achieve the node embeddings of temporal HINs with consideration of not only temporal dynamics but also heterogeneous semantics. The problem is formally described as:

\begin{myproblem}[Temporal HIN embedding]  \label{pro::preli:tHINE}
Given a temporal HIN $\mathcal{G}=(\mathcal{V}, \mathcal{E}, \phi, \varphi, \tau)$, the output is a node representation matrix $\mathbf{X} \in \mathbb{R}^{|\mathcal{V}| \times d}$. Each row of $\mathbf{X}$ is an embedding vector that corresponds to a node and $d \ll |\mathcal{V}|$ is the number of embedding dimensions. The representation matrix $\mathbf{X}$ needs to keep the influence of edge timestamps and node/edge types.
\end{myproblem}

\subsection{Hyperbolic Geometry}

\begin{figure}[t]
    \centering
    \subfigure[\scriptsize{Distance in the Poincaré ball}]{
        \begin{minipage}[t]{0.38\textwidth}
            \centering
            \includegraphics[width=1\textwidth]{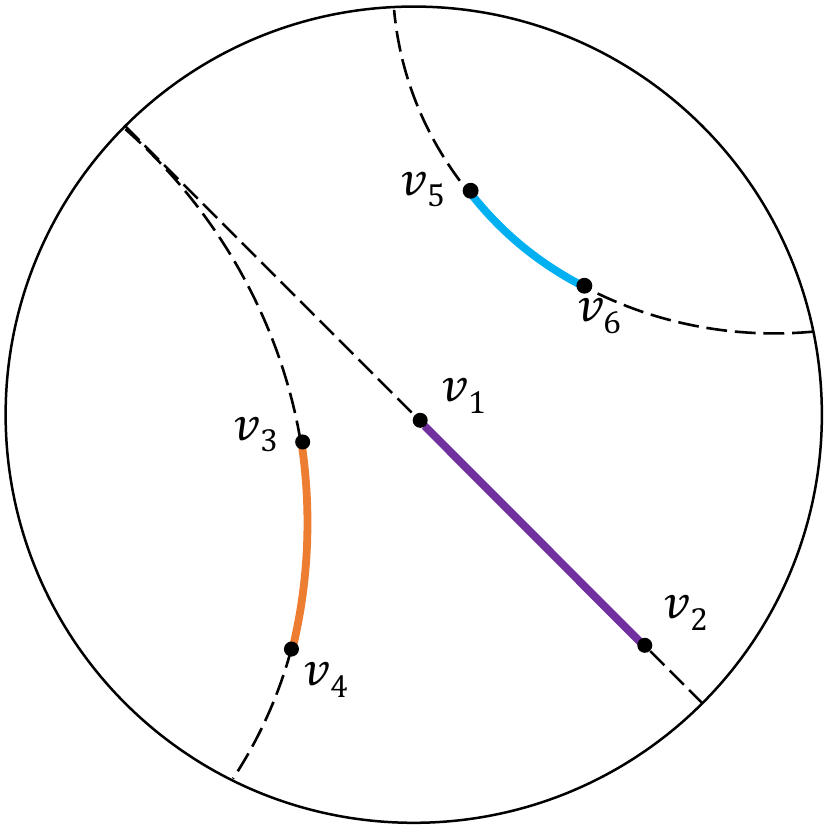}
        \end{minipage}
        \label{fig::preli:hyper:poincare}
    }
    \hspace{5mm}
    \subfigure[\scriptsize{Embedding a tree in $\mathbb{D}$}]{
        \begin{minipage}[t]{0.38\textwidth}
            \centering
            \includegraphics[width=1\textwidth]{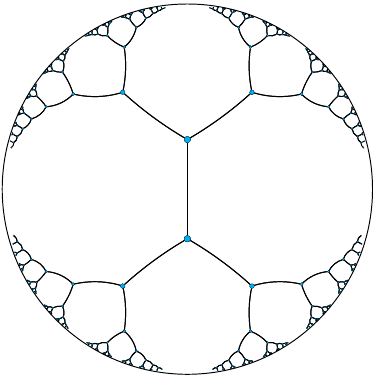}
        \end{minipage}
        \label{fig::preli:hyper:tree_example}
    }
    \caption{(a) The colored curvatures $\mathop{v_1v_2}\limits^{\frown}$, $\mathop{v_3v_4}\limits^{\frown}$ and $\mathop{v_5v_6}\limits^{\frown}$ indicate the distances between nodes in the Poincaré ball. Geodesics are drawn as dashed curvatures. It's worth noting that the distance increases exponentially relative to Euclidean distances with the points being closer to the boundary. Best viewed in color. (b) Referred from \cite{nickelPoincareEmbeddingsLearning}. It shows an example that a tree (a network with power-law distribution) is embedded into the Poincaré ball.}
    \label{fig::preli:hyper}
\end{figure}

Next, we introduce hyperbolic spaces. The $n$-dimensional hyperbolic space $\mathbb{H}^n$ is the unique simply connected $n$-dimensional complete Riemannian manifold with a constant negative sectional curvature. One critical property of hyperbolic spaces is that they expand exponentially, which means the areas of disks with radius $r$ are of $O(e^r)$. This leads to the conclusion that data with power-law distribution is natural to be modeled in hyperbolic spaces~\cite{krioukovHyperbolicGeometryComplex2010}. Although hyperbolic spaces are not isometric to Euclidean spaces and difficult to perform operations on them consequently, there exist several well-known equivalent models of hyperbolic spaces defined on different Euclidean domains, such as the Klein model, the Poincaré ball model and the half-plane model.
% \textcolor{blue}{[Although it is difficult to envisage on hyperbolic spaces because they are not isometric to Euclidean spaces, there has been several well-known equivalent models of hyperbolic spaces defined on different Euclidean domains, such as the Klein model, the Poincaré ball model and the half-plane model. Among them, the Poincaré ball model]} 
The Poincaré ball model is widely used because it is suitable for gradient-based optimization~\cite{nickelPoincareEmbeddingsLearning}. For an $n$-dimensional hyperbolic space with the curvature $c$, the definition domain of corresponding Poincaré ball model $\mathbb{D}$ is the point set
\begin{equation}
    \mathbb{D}=\left\{(x_1, x_2, \dots, x_n): \sum_{i=1}^{n}x_i^2 < -\frac{1}{c}\right\}.
    \label{eq::preli:defdomain}
\end{equation}
In this paper, let $c=-1$ if there are no special instructions. Under this circumstance, the Poincaré ball becomes an open unit ball. The distance between two points $\mathbf{u}$ and $\mathbf{v}$ in the ball is
\begin{equation} 
    d_{\mathbb{D}}(\mathbf{u}, \mathbf{v})=arcosh\left(1+\frac{2\|\mathbf{u}-\mathbf{v}\|^2}{(1-\|\mathbf{u}\|^2)(1-\|\mathbf{v}\|^2)}\right),
    \label{eq::preli:distance}
\end{equation}
where $arcosh(x)=\ln (x+\sqrt{x^2-1})$ is the inverse hyperbolic cosine function. 
Note that the variation of distance is influenced by the location of $\mathbf{u}$ and $\mathbf{v}$. When $(1-\|\mathbf{u}\|^2) \rightarrow 0$ and $(1-\|\mathbf{v}\|^2) \rightarrow 0$, the points are close to the boundary of Poincaré ball and the distance between them is much larger than the case that they are closer to the center. See Figure~\ref{fig::preli:hyper:poincare} for an illustration and Figure~\ref{fig::preli:hyper:tree_example} gives an example for embedding a tree-like network into the Poincaré ball.

% \zhhw{Note that the variation of distance is influenced by the location of $\mathbf{u}$ and $\mathbf{v}$. For two points $\mathbf{u}$ and $\mathbf{v}$, $(1-\|\mathbf{u}\|^2) \rightarrow 0$ and $(1-\|\mathbf{v}\|^2) \rightarrow 0$ when they are close to the boundary of the Poincaré ball. So the distance between them is much larger than the case that they are closer to the center.
% }

\section{Proposed Model}  \label{sec::model}
In this section, we describe our proposed \hhtne ~in details. As Figure~\ref{fig::model:random_walk} shows, \hhtne\\ first leverages a temporally and heterogeneously double-constrained random walk strategy~\cite{jiawenguoJiYuFeiDiJianShiXuSuiJiYouZouDeDongTaiYiZhiWangLuoQianRu2021} to capture both the topological structure and contextual semantics over time. Then, we propose a hyperbolic model defined in Poincaré ball to calculate the embeddings of nodes. Furthermore, we introduce how to optimize the model and analyze the time and space complexity.

\subsection{The Double-constrained Random Walk}  \label{sec::model:randomwalk}
In temporal networks, random walk over time is a natural choice. But in real world, one event may cause several additions of edges, and these edges share the same timestamp. This means local structures with the same timestamp always imply strong relations and semantics. Therefore, within the random walk going over time, a reasonable strategy needs to allow the next hop to stay at current timestamp. So how to make the trade-off between time going and staying becomes a challenge.

% \zhhw{In real-life temporal networks, random walks always change over time because one event may cause several updates of related edges, which share the same timestamp. As a result, local structures with the same timestamp always imply strong relations and semantics. In order to capture both structural features and time features, how to  trade-off current timestamp and next timestamp becomes a challenge. On the one hand, time features can be obtained by modeling the ongoing timestamps. On the other hand, staying at some certain timestamps could extract more structural features. [\textcolor{blue}{I don't know whether this paragraph modified by me could represent your opinion.}]}

Non-decreasing temporal random walk is to run random walk process on a given temporal network $\mathcal{G}$ with the non-decreasing order on timestamps. Each generated node sequence $(v_1, v_2, \dots, v_l)$ conforms to the following rules:
\begin{enumerate}
\item For $i=2, \dots, l-1$, the timestamps between adjacent edges obey $\tau(v_{i-1}, v_i) \leq \tau(v_i, v_{i+1})$.
\item The probability of $\tau(v_{i-1}, v_i)=\tau(v_i, v_{i+1})$, which means the timestamp of next hop equals to current timestamp, descends with the increase of hops staying at the same timestamp.
\end{enumerate}
The latter tries to ensure that the random walk goes over time, and meanwhile keeps the possibility of staying at current timestamp. Let $t_i=\tau(v_{i-1}, v_i)$, then the probability of next hop staying at $t_i$ is:
\begin{equation}
    P\left( \tau(v_i, v_{i+1})=t_i \right) = \left\{ \begin{array}{ll} \text{stop} & \text{if} ~~ \mathcal{N}_{v_i}(t_i^*)=\varnothing, \\ 0 & \text{if} ~~ t_i \notin \mathcal{T}(\mathcal{N}_{v_i}), \\ \beta^n & \text{otherwise}.  \end{array} \right.
    \label{eq::model:temporal}
\end{equation}
Here, $\mathcal{N}_v$ is the universal neighbor set of node $v$,  $\mathcal{N}_v(t^*)$ is the set of those connected with node $v$ at timestamp $t$ or later, $\mathcal{T}(\mathcal{N}_v)$ denotes the timestamp set of edges between node $v$ and its neighbors. $\beta \in [0, 1]$ is the initial timestamp staying probability and $n$ refers to the number of hops for which the random walk have been at current timestamp. First, in case that all the edges connected to node $v_i$ are before timestamp $t_i$, we can only stop this random walk. Second, in case that there are no edges connected to node $v_i$ at timestamp $t_i$, the next hop can only be chosen from the edges after $t_i$. Finally, if the timestamps of edges between node $v_i$ and its neighbors contain both $t_i$ and those after $t_i$, we probabilistically control the next hop staying at timestamp $t_i$ with a probability $\beta^n$, and going to later timestamps otherwise. Adopting an exponential decay function here penalizes the cases that the random walk stays at the same timestamp for too long. 

\begin{figure}[t]
    \centering
    \includegraphics[width=0.65\textwidth]{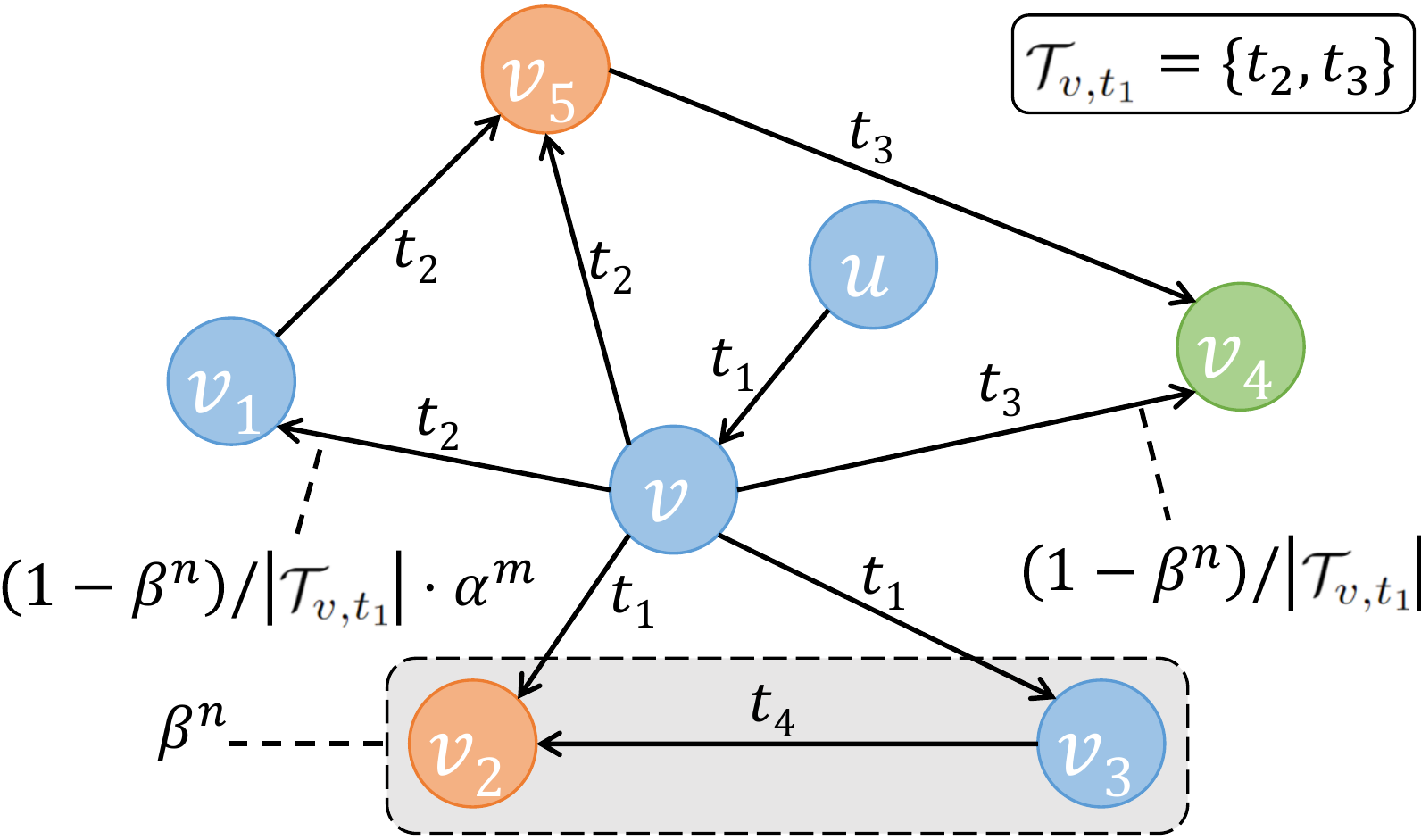}
    \caption{The temporally and heterogeneously double-constrained random walk. Different colors refer to different node types and the timestamps are labeled on the edges. Last hop of the random walk is from node $u$ to node $v$ at timestamp $t_1$. Best viewed in color.}
    \label{fig::model:random_walk}
\end{figure}

Taking Figure~\ref{fig::model:random_walk} as an example, suppose that the random walk jumps from node $u$ to $v$ at timestamp $t_1$ in the last hop and it has been at $t_1$ for $n$ hops. The probability of next hop staying at $t_1$, which means jumping to $v_2$ or $v_3$, ~is $\beta^n$. Let $\mathcal{T}_{v, t_1}$ denote the set of timestamps related to $v$ after $t_1$. $\forall t \in \mathcal{T}_{v, t_1}$, the probability of the timestamp of next hop being $t$ is $(1-\beta^n)/|\mathcal{T}_{v, t_1}|$.

The random walks on temporal HINs need to satisfy two constraints: one is the non-decreasing principle for temporal information, and the other is for heterogeneous information. Taking node types into account, most existing methods use metapaths to guide the random walk sampling. However, in our model, overlaying the temporal constraint, metapaths are too strict to ensure the random walk process going continuously. Inspired by \cite{husseinAreMetaPathsNecessary2018a}, we leverage a biased type choosing strategy based on historical walks. Suppose that we have chosen $t_i$ to be the timestamp of next hop and currently the random walk is staying at node $v_i$ of the node type $l_i$, then the probability of next hop still staying at node type $l_i$ is:
\begin{equation}
    P\left(\phi(v_{i+1})=l_i ; t_i \right)=\left \{ \begin{array}{ll} 0 & \text{if} ~~ \mathcal{N}_{v_i}(t_i) \cap \mathcal{N}_{v_i}(l_i)=\varnothing, \\ 1 & \text{if} ~~ \mathcal{N}_{v_i}(t_i) \subseteq \mathcal{N}_{v_i}(l_i), \\ \alpha^m & \text{otherwise}. \end{array} \right.
    \label{eq::model:heterogeneous}
\end{equation}
$\mathcal{N}_v(t)$ denotes the set of neighbors which connect to node $v$ exactly at timestamp $t$, and $\mathcal{N}_v(l)$ refers to neighbors of the node type $l$. $\alpha \in [0, 1]$ is the initial node type staying probability and $m$ is the number of hops for which the random walk have been at node type $l$. Similar to the timestamp of next hop as aforementioned, we consider the following three cases. If any neighbors that connect to $v_i$ at timestamp $t_i$ are not of type $l_i$, the next hop will surely go to another node type. If all neighbors of $v_i$ at timestamp $t_i$ are of type $l_i$, the random walk can only stay at the same node type. If part of neighbors at timestamp $t_i$ are of type $l_i$ while others are not, the probability of next hop being at type $l_i$ should be $\alpha^m$. Similar to Eq.~\ref{eq::model:temporal}, we adopt the second exponential decay function to avoid the random walk staying at the same node type continuously.

Back to Figure~\ref{fig::model:random_walk}, suppose that the random walk has been staying at the same node type (in blue) for $m$ hops. If we have chosen $t_2$ as the timestamp of next hop, according to Eq.~\ref{eq::model:heterogeneous}, the probability of staying at current type is $\alpha^m$. Overlaying the temporal constraint, the random walk will jump to node $v_1$ in next hop with the probability $(1-\beta^n)/|\mathcal{T}_{v, t_1}| \cdot \alpha^m$.

With above constraints, we incrementally update the random walks when new edges come in order to lower the complexity of our model. In addition to newly added edges, historical random walks should also be processed appropriately. On the one hand, in real world, relations between entities are effected by historical events so they need to be preserved. On the other hand, their influence decreases gradually over time and recent events have greater importance for relations among entities.

The evolution of networks is manifested in the addition and deletion of nodes and edges. When evolution happens, nodes which directly connected to new nodes and edges are involved, and the random walks should be updated correspondingly. We consider the following four cases to incrementally update the node sequences of random walks when new edges come:
\begin{enumerate}
\item \textbf{Preserve uninvolved sequences.} If none of nodes in a sequence is involved, we keep the sequence unchanged.
\item \textbf{Remove invalid parts of sequences.} If the timestamps of front part of a sequence are too early, we treat this part as invalid and remove it.
\item \textbf{Continue involved sequences.} If the the last node of a sequence is involved, continue the random walk with above strategy.
\item \textbf{Reverse new random walks from new timestamps.} When new edges and nodes come, to ensure that these changes are considered, reversed random walks from new timestamps backwards are processed.
\end{enumerate}

With all above, we have captured the temporal and heterogeneous information of the networks and imply it into node sequences of random walks. Next, we'll explain how to embed these nodes into the hyperbolic space.

\subsection{Hyperbolic Embedding}  \label{sec::model:embed}
Defined in Poincaré ball, we exploit the hyperbolic distance in Eq.~\ref{eq::preli:distance} to measure the proximity between nodes and calculate their probability of co-occurrence in random walks as following:
\begin{equation}
    P(u|v) = \sigma \left( -d_{\mathbb{D}} (\textbf{u}, \textbf{v}) \right),
    \label{eq::model:neighbor_judge}
\end{equation}
where $\sigma(z)=\frac{1}{1+e^{-z}}$, $u|v$ denotes the co-occurrence of node $u$ and $v$, and $\textbf{u}$ and $\textbf{v}$ denote the embedding vector of node $u$ and $v$. For each node $u$, we randomly sample $k$ negative nodes, each of which is denoted by $n$ and each of whose corresponding embedding vector is denoted by $\textbf{n}$, to speed up the training. The proximity between $u$ and the co-occurrent node  $v$ is expected to be higher, so the optimization goal for each node $u$ is:
\begin{equation}
    \arg\max_{\Theta} ~ \log \sigma \left( -d_{\mathbb{D}} (\textbf{u}, \textbf{v}) \right) + \sum_n \log \sigma \left( d_{\mathbb{D}} (\textbf{u}, \textbf{n}) \right).
    \label{eq::model:object_node_neg}
\end{equation}

Summing up all nodes and enhancing the ranking, the objective function of \hhtne~ is written as:
\begin{equation}
\begin{split}
    \mathcal{L} &= \arg\max_{\Theta} \sum_{u \in \mathcal{N}} \sum_{v; u|v} \log \sigma \left( -d_{\mathbb{D}} (\textbf{u}, \textbf{v}) \right) + \sum_{u \in \mathcal{N}} \sum_{n;neg(u)} \log \sigma \left( d_{\mathbb{D}} (\textbf{u}, \textbf{n}) \right) \\
    & \sim \arg\max_{\Theta} \sum_{u \in \mathcal{N}} \sum_{v; u|v} \sum_{n;neg(u)} \log \sigma \left( d_{\mathbb{D}} (\textbf{u}, \textbf{n}) - d_{\mathbb{D}} (\textbf{u}, \textbf{v}) \right),
    \label{eq::model:object}
\end{split}
\end{equation}
where $neg(u)$ denotes the negative samples for node $u$.

\vspace{0.5em}
\noindent
\textbf{Optimization.} Due to the Riemannian manifold structure of Poincaré ball, the optimization is different from Euclidean models. Following \cite{nickelPoincareEmbeddingsLearning}, we primarily calculate the Euclidean gradients.

In training process, the parameters $\Theta$ are updated for each step of calculation. For $l=\log \sigma \left( d_{\mathbb{D}} (\textbf{u}, \textbf{n}) - d_{\mathbb{D}} (\textbf{u}, \textbf{v}) \right)$, 
\begin{equation}
\begin{split}
    \frac{\partial l}{\partial \textbf{u}} & = \left( 1 - \sigma \left( d_{\mathbb{D}} (\textbf{u}, \textbf{n}) - d_{\mathbb{D}} (\textbf{u}, \textbf{v}) \right) \right) \cdot \left( \frac{\partial d_{\mathbb{D}} (\textbf{u}, \textbf{n})}{\partial \textbf{u}}  - \frac{\partial d_{\mathbb{D}} (\textbf{u}, \textbf{v})}{\partial \textbf{u}} \right),\\
    \frac{\partial l}{\partial \textbf{v}} & = \left( 1 - \sigma \left( d_{\mathbb{D}} (\textbf{u}, \textbf{n}) - d_{\mathbb{D}} (\textbf{u}, \textbf{v}) \right) \right) \cdot \left( - \frac{\partial d_{\mathbb{D}} (\textbf{u}, \textbf{v})}{\partial \textbf{v}} \right),\\
    \frac{\partial l}{\partial \textbf{n}} & = \left( 1 - \sigma \left( d_{\mathbb{D}} (\textbf{u}, \textbf{n}) - d_{\mathbb{D}} (\textbf{u}, \textbf{v}) \right) \right) \cdot \left( \frac{\partial d_{\mathbb{D}} (\textbf{u}, \textbf{n})}{\partial \textbf{n}} \right).
    \label{eq::model:grad_step}
\end{split}
\end{equation}
$\frac{\partial d_{\mathbb{D}} (\textbf{u}, \textbf{v})}{\partial \textbf{u}}$, $\frac{\partial d_{\mathbb{D}} (\textbf{u}, \textbf{v})}{\partial \textbf{v}}$, $\frac{\partial d_{\mathbb{D}} (\textbf{u}, \textbf{n})}{\partial \textbf{u}}$ and $\frac{\partial d_{\mathbb{D}} (\textbf{u}, \textbf{n})}{\partial \textbf{n}}$ is further derived:
\begin{equation}
\begin{split}
    \frac{\partial d_{\mathbb{D}} (\textbf{u}, \textbf{v})}{\partial \textbf{u}} & = \frac{4}{\delta_v \sqrt{\gamma_{uv}^2-1}} \left( \frac{\|\textbf{v}\|^2 - 2 \langle \textbf{u}, \textbf{v} \rangle + 1}{\delta_u^2}\textbf{u} - \frac{\textbf{v}}{\delta_u} \right), \\
    \frac{\partial d_{\mathbb{D}} (\textbf{u}, \textbf{v})}{\partial \textbf{v}} & = \frac{4}{\delta_u \sqrt{\gamma_{uv}^2-1}} \left( \frac{\|\textbf{u}\|^2 - 2 \langle \textbf{u}, \textbf{v} \rangle + 1}{\delta_v^2}\textbf{v} - \frac{\textbf{u}}{\delta_v} \right), \\
    \frac{\partial d_{\mathbb{D}} (\textbf{u}, \textbf{n})}{\partial \textbf{u}} & = \frac{4}{\delta_n \sqrt{\gamma_{un}^2-1}} \left( \frac{\|\textbf{n}\|^2 - 2 \langle \textbf{u}, \textbf{n} \rangle + 1}{\delta_u^2}\textbf{u} - \frac{\textbf{n}}{\delta_u} \right), \\
    \frac{\partial d_{\mathbb{D}} (\textbf{u}, \textbf{n})}{\partial \textbf{n}} & = \frac{4}{\delta_u \sqrt{\gamma_{un}^2-1}} \left( \frac{\|\textbf{u}\|^2 - 2 \langle \textbf{u}, \textbf{n} \rangle + 1}{\delta_n^2}\textbf{n} - \frac{\textbf{u}}{\delta_n} \right),
    \label{eq::model:grad_dis}
\end{split}
\end{equation}
where $\delta_u=1-\|\textbf{u}\|^2$, $\delta_v=1-\|\textbf{v}\|^2$, $\delta_n=1-\|\textbf{n}\|^2$, $\gamma_{uv}=1+\frac{2}{\delta_u \delta_v}\|\textbf{u}-\textbf{v}\|^2$ and $\gamma_{un}=1+\frac{2}{\delta_u \delta_n}\|\textbf{u}-\textbf{n}\|^2$. 

Next, combining with Riemannian gradient, a single embedding is updated as follows:
\begin{equation}
\begin{split}
    \textbf{u}_{new} & \leftarrow \text{proj} \left( \textbf{u}_{old} + lr \frac{(1-\|\textbf{u}_{old}\|^2)^2}{4} \frac{\partial l}{\partial \textbf{u}} \right), \\
    \textbf{v}_{new} & \leftarrow \text{proj} \left( \textbf{v}_{old} + lr \frac{(1-\|\textbf{v}_{old}\|^2)^2}{4} \frac{\partial l}{\partial \textbf{v}} \right), \\
    \textbf{n}_{new} & \leftarrow \text{proj} \left( \textbf{n}_{old} + lr \frac{(1-\|\textbf{n}_{old}\|^2)^2}{4} \frac{\partial l}{\partial \textbf{n}} \right), 
    \label{eq::model:update}
\end{split}
\end{equation}
in which $lr$ is the learning rate and $\text{proj}(\cdot)$ is a projection function that constrains the embeddings within the Poincaré ball:
\begin{equation}
    \text{proj}(\textbf{x}) = \left\{ \begin{array}{ll} \textbf{x} / (\|\textbf{x}\|+\epsilon) ~~~~ & \text{if} \|\textbf{x}\| \geq 1, \\
    \textbf{x} & \text{otherwise}.
    \end{array} \right.
    \label{eq::model:proj}
\end{equation}
Here $\epsilon=10^{-7}$ is a small constant.

\subsection*{Time Complexity Analysis}  \label{sec::model:complexity}
The time complexity for random walk updating is $\mathcal{O}(|\mathcal{N}| \cdot \Delta_{l_{avg}})$, in which $\Delta_{l_{avg}}$ is the average length of updated part in random walks. The time complexity of hyperbolic embedding training is $\mathcal{O}(I \cdot k \cdot d \cdot |\mathcal{N}| \cdot \Delta_{l_{avg}})$, where $I$ is the number of iterations, $k$ is the number of negative samples for each node, and $d$ is the dimension number of embedding space.

\section{Experiments and Discussions}  \label{sec::expr}

\subsection{Experimental Setup}  \label{sec::expr:setup}
% \vspace{0.3em}
% \noindent
\subsubsection{Datasets.} 

\begin{table}[tbp!]
\centering
\setlength{\tabcolsep}{12pt}
\renewcommand{\arraystretch}{1.05}
\caption{Statistics of the datasets.}
\resizebox{0.85\textwidth}{!}{
    \begin{tabular}{c|cccc}
    \toprule
    Datasets & Enron & DBLP & Tokyo & MovieLens \\
    \midrule
    \# Nodes & 115 & 164,174 & 64,151 & 9,940 \\
    \# Edges & 43,160 & 845,485 & 573,703 & 1,000,209 \\
    \# Node types & 9 & 2 & 2 & 2 \\
    \# Timestamps & 20 & 46 & 555,437 & 25,865 \\
    \bottomrule
    \end{tabular}
}
\label{tab::expr:datasets}
\end{table}

We make extensive experiments on four real-world datasets Enron\footnote{\url{http://www.ahschulz.de/enron-email-data/}}, DBLP\footnote{\url{https://www.aminer.cn/citation}}, Tokyo~\cite{dingqiyangModelingUserActivity2015} and MovieLens~\cite{harperMovieLensDatasetsHistory2016}. The statistics of these datasets are shown in Table~\ref{tab::expr:datasets}.

% \vspace{0.5em}
% \noindent
% \subsubsection{Baselines.} 
% Table~\ref{tab::expr:baselines} lists eleven SOTA network embedding methods, including six shallow models and five deep ones.

\begin{table}[tbp!]
\centering
\setlength{\tabcolsep}{14pt}
\renewcommand{\arraystretch}{1.05}
\caption{Baselines.}
\resizebox{0.8\textwidth}{!}{
    \begin{tabular}{c|c|c|c}
    \toprule
    \multicolumn{2}{c|}{Methods} & Heterogeneous & Temporal \\
    \midrule
    \multirow{6}{*}{Shallow} & Deepwalk~\cite{perozziDeepWalkOnlineLearning2014} & $\times$ & $\times$ \\
    & CTDNE~\cite{leeDynamicNodeEmbeddings2020} & $\times$ & $\surd$ \\
    & ISGNS~\cite{pengDynamicNetworkEmbedding2020} & $\times$ & $\surd$ \\
    & Change2vec~\cite{bianNetworkEmbeddingChange2019a} & $\surd$ & $\surd$ \\
    & DHNE~\cite{yinDHNENetworkRepresentation2019} & $\surd$ & $\surd$ \\
    & HHNE~\cite{wangHyperbolicHeterogeneousInformation2019} & $\surd$ & $\times$ \\
    \midrule
    \multirow{5}{*}{Deep} & GCN~\cite{kipfSemiSupervisedClassificationGraph2017} & $\times$ & $\times$ \\
    & GAT~\cite{velickovicGraphAttentionNetworks2018} & $\times$ & $\times$ \\
    & DySAT~\cite{sankarDySATDeepNeural2020} & $\times$ & $\surd$ \\
    & LIME~\cite{pengLIMELowCostIncremental2021} & $\surd$ & $\surd$ \\
    & TGAT~\cite{xuInductiveRepresentationLearning2020} & $\times$ & $\surd$ \\
    \bottomrule
    \end{tabular}
}
\label{tab::expr:baselines}
\end{table}

% \vspace{0.5em}
% \noindent
\subsubsection{Baselines and Experimental Details.} 

Table~\ref{tab::expr:baselines} lists eleven SOTA network embedding methods, including six shallow models and five deep ones.

For all baselines, we take the recommended parameter settings except that the embedding size is set to be 128. For \hhtne, we set the number of walks per node as 10, the maximum of walk length as 80, the number of negative samples per node as 5 and initial learning rate as 0.001. For the experimental results in Section~\ref{sec::expr:results}, we set the values of $\alpha$ and $\beta$ to be 0.9 and 0.3, and our method is trained on the dimensions both 16 and 128, denoted by \hhtne$_{16}$ and \hhtne$_{128}$, respectively. 

In order to validate the effectiveness of each part in our model, we further conduct ablation experiments on three different \hhtne ~variants, in which \hhtne$_{-he}$ denotes ignoring heterogeneous information in random walk process, \hhtne$_{-te}$ denotes ignoring temporal information, and \hhtne$_{-hy}$ denotes embedding in Euclidean spaces. For fairness all of these three variants are evaluated with the dimension of 128.
% \zhhw{You'd better explain the value of hyper parameters for \hhtne if possible.}

We evaluate the performance of \hhtne ~ for temporal link prediction and node classification tasks on a server with 2 $\times$ Intel Xeon Gold 6226R 16C 2.90GHz CPUs, 4 $\times$ GeForce RTX 3090 GPUs and 256 GB memory. The experiments are run on Ubuntu 18.04 with CUDA 11.1.

\subsection{Experimental Performance}  \label{sec::expr:results}

\begin{table}[htbp!]
\centering
\setlength{\tabcolsep}{2pt}
\renewcommand{\arraystretch}{1.05}
\caption{Performance on temporal link prediction.}
\resizebox{\textwidth}{!}{
    \begin{tabular}{c|cccc|cccccc}
    \toprule
    Dataset & \multicolumn{4}{c|}{Enron} & \multicolumn{6}{c}{Tokyo} \\
    \midrule
    Timestamp & 10 & 15 & 20 & Avg. & 200k & 300k & 400k & 500k & 555k & Avg. \\
    \midrule
    Deepwalk & 0.9105 & 0.9199 & 0.9015 & 0.9106 & 0.7854 & 0.8695 & 0.8386 & 0.8565 & 0.8831 & 0.8466 \\
    CTDNE & 0.8403 & \underline{0.9269} & 0.9120 & 0.8931 & 0.7124 & 0.7621 & 0.7519 & 0.7695 & 0.8239 & 0.7640 \\
    ISGNS & \underline{0.9162} & 0.7392 & 0.7384 & 0.7979 & 0.7159 & 0.7315 & 0.6997 & 0.6706 & 0.5779 & 0.6791 \\
    Change2vec & 0.7750 & 0.7796 & 0.7845 & 0.7797 & 0.5614 & 0.5266 & 0.5516 & 0.5164 & 0.4959 & 0.5304 \\
    DHNE & - & 0.5751 & 0.5774 & 0.5763 & - & 0.5057 & 0.5005 & 0.5004 & 0.5002 & 0.5017 \\
    HHNE & 0.8928 & 0.9135 & 0.8992 & 0.9018 & 0.6241 & 0.7225 & 0.7802 & 0.8513 & 0.8770 & 0.7710 \\
    \midrule
    GCN & 0.8847 & 0.8916 & 0.8547 & 0.877 & 0.4944 & 0.4052 & 0.3626 & 0.3078 & 0.3004 & 0.3741 \\
    GAT & \textbf{0.9360} & 0.8867 & \textbf{0.9292} & \underline{0.9173} & 0.6911 & 0.6690 & 0.6518 & 0.7328 & 0.7086 & 0.6907 \\
    DySAT & - & 0.7456 & 0.8305 & 0.7881 & - & 0.6576 & 0.6462 & 0.6473 & 0.6504 & 0.6504 \\
    LIME & 0.5642 & 0.6296 & 0.5116 & 0.5685 & 0.5166 & 0.5144 & 0.5141 & 0.5107 & 0.5125 & 0.5137 \\
    TGAT & 0.7224 & 0.6814 & 0.6799 & 0.6946 & 0.6432 & 0.6731 & 0.4647 & 0.6593 & 0.7365 & 0.6353 \\
    \midrule
    \hhtne$_{-he}$ & 0.7886 & 0.9144 & 0.8994 & 0.8675 & 0.8753 & 0.8953 & 0.8848 & 0.8855 & 0.9008 & \underline{0.8883} \\
    \hhtne$_{-te}$ & 0.7172 & 0.8954 & 0.9044 & 0.8390 & 0.8773 & 0.8936 & 0.8839 & 0.8834 & 0.8980 & 0.8872 \\
    \hhtne$_{-hy}$ & 0.9153 & \textbf{0.9363} & \underline{0.9204} & \textbf{0.9240} & 0.7582 & 0.7907 & 0.7615 & 0.7779 & 0.8226 & 0.7822 \\
    \midrule
    \hhtne$_{16}$ & 0.9052 & 0.9199 & 0.9060 & 0.9104 & \underline{0.8432} & \underline{0.8795} & \underline{0.8702} & \underline{0.8727} & \underline{0.8921} & 0.8715 \\
    \hhtne$_{128}$ & 0.9106 & 0.9233 & 0.9105 & 0.9148 & \textbf{0.8795} & \textbf{0.8977} & \textbf{0.8871} & \textbf{0.8872} & \textbf{0.9032} & \textbf{0.8910} \\
    \bottomrule
    \end{tabular}
}
\resizebox{\textwidth}{!}{
    \begin{tabular}{c|ccccc|ccccc}
    \toprule
    Dataset & \multicolumn{5}{c|}{DBLP} & \multicolumn{5}{c}{MovieLens} \\
    \midrule
    Timestamp & 20 & 30 & 40 & 46 & Avg. & 10k & 15k & 20k & 25k & Avg. \\
    \midrule
    Deepwalk & 0.9290 & 0.8710 & 0.8422 & 0.8378 & 0.8700 & 0.8816 & 0.8482 & 0.8607 & 0.8310 & 0.8554 \\
    CTDNE & 0.7381 & 0.5199 & 0.5657 & 0.5424 & 0.5915 & 0.3248 & 0.4165 & 0.4551 & 0.4577 & 0.4135 \\
    ISGNS & \underline{0.9321} & 0.8579 & 0.8414 & 0.7788 & 0.8526 & 0.8355 & 0.8030 & 0.8070 & 0.7771 & 0.8057 \\
    Change2vec & 0.5482 & 0.5161 & 0.5025 & 0.4989 & 0.5164 & 0.8438 & 0.8455 & 0.8643 & 0.8446 & 0.8496 \\
    DHNE & - & 0.5177 & 0.5002 & 0.5002 & 0.5060 & - & 0.5007 & 0.5004 & 0.4995 & 0.5002 \\
    HHNE & 0.3487 & 0.5956 & 0.6987 & 0.7195 & 0.5906 & 0.8683 & 0.8881 & 0.8834 & 0.8837 & 0.8809 \\
    \midrule
    GCN & 0.6550 & 0.6308 & 0.5706 & 0.5878 & 0.6111 & 0.2318 & 0.2343 & 0.2491 & 0.1816 & 0.2242 \\
    GAT & 0.6722 & 0.7472 & 0.6474 & 0.5657 & 0.6581 & 0.4960 & 0.2448 & 0.3311 & 0.3889 & 0.3652 \\
    DySAT & - & 0.5178 & 0.6046 & 0.6272 & 0.5832 & - & 0.7463 & 0.7429 & 0.7167 & 0.7353 \\
    LIME & 0.5041 & 0.5353 & 0.5392 & 0.5385 & 0.5293 & 0.4801 & 0.4753 & 0.4764 & 0.4786 & 0.4776 \\
    TGAT & 0.5091 & 0.6215 & 0.3311 & 0.5648 & 0.5066 & 0.4000 & 0.4567 & 0.5326 & 0.5161 & 0.4764 \\
    \midrule
    \hhtne$_{-he}$ & 0.7270 & 0.8151 & 0.8376 & 0.8379 & 0.8044 & 0.8899 & 0.9276 & 0.9450 & 0.9292 & 0.9229 \\
    \hhtne$_{-te}$ & 0.7681 & 0.6947 & 0.7849 & 0.8007 & 0.7621 & 0.6925 & 0.6870 & 0.6987 & 0.7156 & 0.6985 \\
    \hhtne$_{-hy}$ & 0.7687 & 0.7989 & 0.7621 & 0.7573 & 0.7718 & 0.7448 & 0.5182 & 0.5385 & 0.5597 & 0.5903 \\
    \midrule
    \hhtne$_{16}$ & 0.9302 & \underline{0.8770} & \underline{0.8757} & \underline{0.8505} & \underline{0.8834} & \underline{0.9264} & \underline{0.9433} & \underline{0.9466} & \underline{0.9377} & \underline{0.9385} \\
    \hhtne$_{128}$ & \textbf{0.9349} & \textbf{0.8812} & \textbf{0.8770} & \textbf{0.8582} & \textbf{0.8878} & \textbf{0.9380} & \textbf{0.9475} & \textbf{0.9555} & \textbf{0.9435} & \textbf{0.9461} \\
    \bottomrule
    \end{tabular}
}
\flushleft{{$^*$ The best result is in bold, and the suboptimal result is underlined. The same applies to below tables.}}
\label{tab::expr:link_prediction}
\end{table}

\begin{table*}[htbp!]
\centering
\setlength{\tabcolsep}{2pt}
\renewcommand{\arraystretch}{1.05}
\caption{Performance on node classification.}
\resizebox{\textwidth}{!}{
    \begin{tabular}{c|cc|cc|cc|cc}
    \toprule
    Dataset & \multicolumn{2}{c|}{Enron} & \multicolumn{2}{c|}{DBLP} & \multicolumn{2}{c|}{Tokyo} & \multicolumn{2}{c}{MovieLens} \\
    \midrule
    Metrics & Macro-f1 & Micro-f1 & Macro-f1 & Micro-f1 & Macro-f1 & Micro-f1 & Macro-f1 & Micro-f1 \\
    \midrule
    Deepwalk & 0.1832 & 0.3793 & 0.4999 & 0.9995 & 0.5208 & 0.9617 & 0.9446 & 0.9531 \\
    CTDNE & 0.1277 & 0.2828 & 0.6552 & 0.9993 & 0.4915 & 0.9666 & 0.9862 & 0.9871 \\
    ISGNS & 0.1084 & 0.3034 & \underline{0.7438} & 0.9995 & 0.8501 & 0.9815 & 0.9903 & 0.9909 \\
    Change2vec & 0.4482 & 0.6414 & 0.4997 & 0.9989 & 0.5229 & 0.9596 & 0.9415 & 0.9628 \\
    DHNE & 0.0641 & 0.3448 & 0.4998 & 0.9994 & 0.4910 & 0.9648 & 0.4002 & 0.6036 \\
    HHNE & 0.1497 & 0.2414 & 0.4997 & 0.9986 & 0.7953 & \underline{0.9889} & 0.9906 & 0.9912 \\
    \midrule
    GCN & 0.2957 & 0.4344 & 0.4999 & 0.9994 & 0.5280 & 0.9644 & \textbf{0.9995} & \textbf{0.9995} \\
    GAT & 0.1824 & 0.2828 & 0.4998 & 0.9993 & 0.5723 & 0.9674 & 0.9599 & 0.9624 \\
    DySAT & 0.0972 & 0.2069 & 0.6358 & 0.9977 & \underline{0.9095} & 0.9816 & 0.9445 & 0.9478 \\
    LIME & 0.0794 & 0.1538 & 0.4998 & 0.9955 & 0.4818 & 0.9299 & 0.8188 & 0.8317 \\
    TGAT & 0.1018 & 0.2069 & 0.4998 & 0.9992 & 0.7381 & 0.9689 & 0.9459 & 0.9496 \\
    \midrule
    \hhtne$_{-he}$ & 0.1241 & 0.3103 & 0.5907 & 0.9992 & 0.4908 & 0.9639 & 0.9866 & 0.9875 \\
    \hhtne$_{-te}$ & 0.4215 & 0.5517 & 0.6199 & 0.9995 & 0.4903 & 0.9618 & 0.9880 & 0.9888 \\
    \hhtne$_{-hy}$ & 0.1628 & 0.3172 & \textbf{0.8023} & \underline{0.9996} & \textbf{0.9469} & \textbf{0.9928} & 0.8434 & 0.8523 \\
    \midrule
    \hhtne$_{16}$ & \textbf{0.8121} & \textbf{0.7931} & 0.4999 & \textbf{0.9997} & 0.4916 & 0.9670 & 0.9873 & 0.9915 \\
    \hhtne$_{128}$ & \underline{0.6783} & \textbf{0.7931} & 0.6499 & \underline{0.9996} & 0.4913 & 0.9659 & \textbf{0.9995} & \textbf{0.9995} \\
    \bottomrule
    \end{tabular}
}
\label{tab::expr:node_classification}
\end{table*}

We train the baselines shown in Table~\ref{tab::expr:baselines} in unsupervised manners on all datasets and evaluate them by two traditional tasks, temporal link prediction and node classification. 

% \vspace{2mm}
% \noindent
% \textbf{Temporal Link Prediction.}
\subsubsection{Temporal Link Prediction.} 
In this task, we divide Enron, DBLP, Tokyo and MovieLens into 4, 5, 6 and 5 snapshots evenly according to timestamps, and for the $t$-th snapshot except the first, the models are trained on the first $t-1$ snapshots and tested on the $t$-th snapshot. Noting that the Poincaré ball is a conformal model, which means that the angles in hyperbolic spaces are equal to corresponding ones in Euclidean spaces, we use the cosine similarity to calculate the proximity between nodes for fairness to all models. AUC is used to measure the performance in this task. The experimental results for all snapshots are shown in Table~\ref{tab::expr:link_prediction}.  In most cases, our method has the best performance and even 16-dimension embeddings of \hhtne~ are better than previous models in dimension 128. On the dense dataset Enron, our method does not outperform other Euclidean models, suggesting that hyperbolic models are more suitable for sparse data, which is consistent with ~\cite{zhangWhereAreWe2021}. In addition, the results of ablation models show that performance of different variants mostly degrades. It proves that double-constrained random walk strategy does capture the temporal and heterogeneous information and the hyperbolic space does enhance the representation ability of our model. 

% \vspace{2mm}
% \noindent
% \textbf{Node Classification.}
\subsubsection{Node Classification.} 
In this task, we use 75\% of node embeddings for each dataset to train a Logistic Regression classifier and the remains are treated as the test set. The results are evaluated by two metrics macro-f1 and micro-f1. As Table~\ref{tab::expr:node_classification} shows, our method outperforms other models in most cases. It's worth noting that for most methods in datasets DBLP and Tokyo, the difference between macro-f1 and micro-f1 is very large, which is caused by the imbalance of various node types.

\subsection{Parameter Analysis}  \label{sec::expr:parameter}

\begin{figure}[tb!]
    \centering
    \subfigure[]{
        \begin{minipage}[t]{0.48\textwidth}
            \centering
            \includegraphics[width=0.95\textwidth,height=0.7\textwidth]{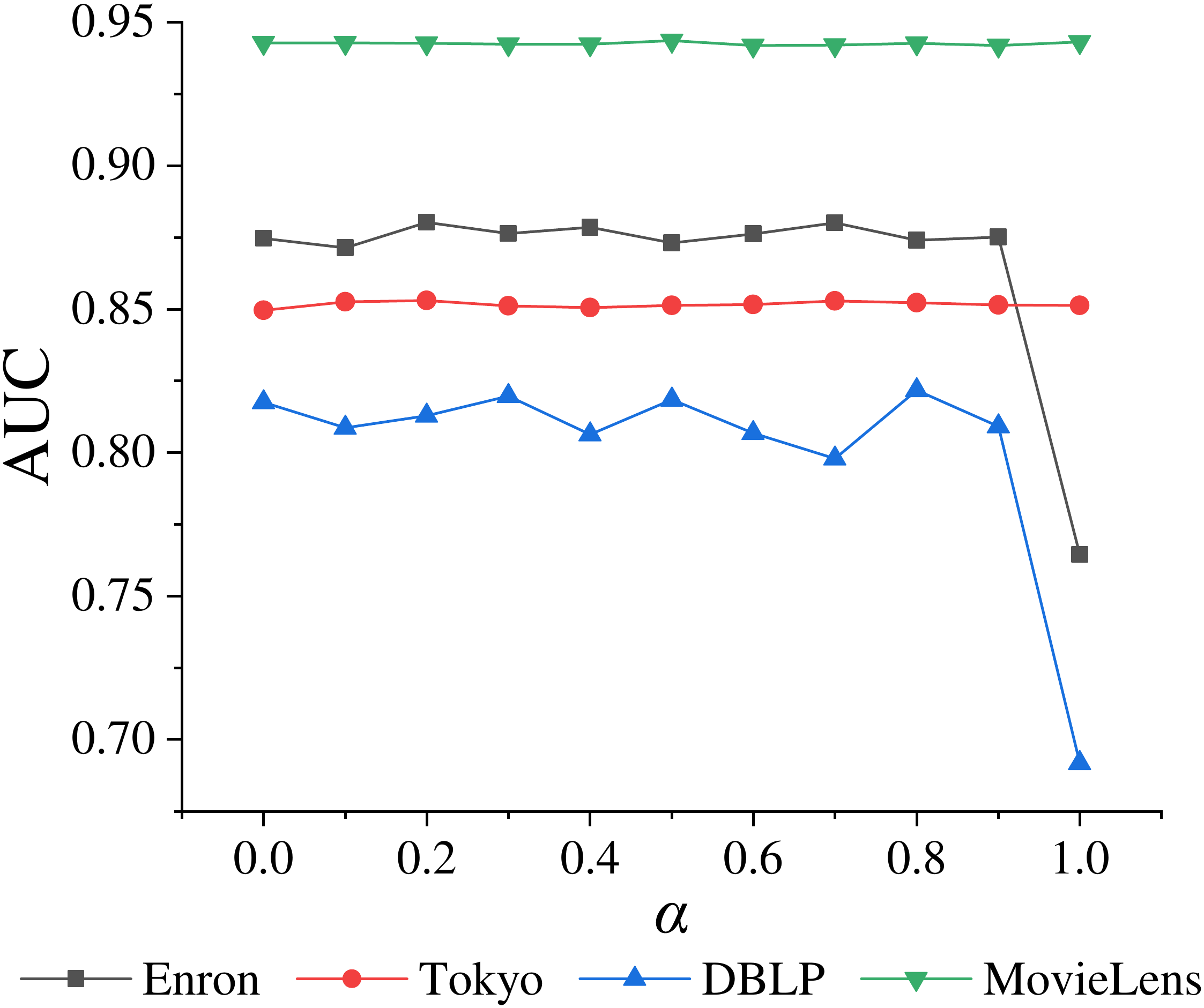}
        \end{minipage}
    }
    \subfigure[]{
        \begin{minipage}[t]{0.48\textwidth}
            \centering
            \includegraphics[width=0.95\textwidth,height=0.7\textwidth]{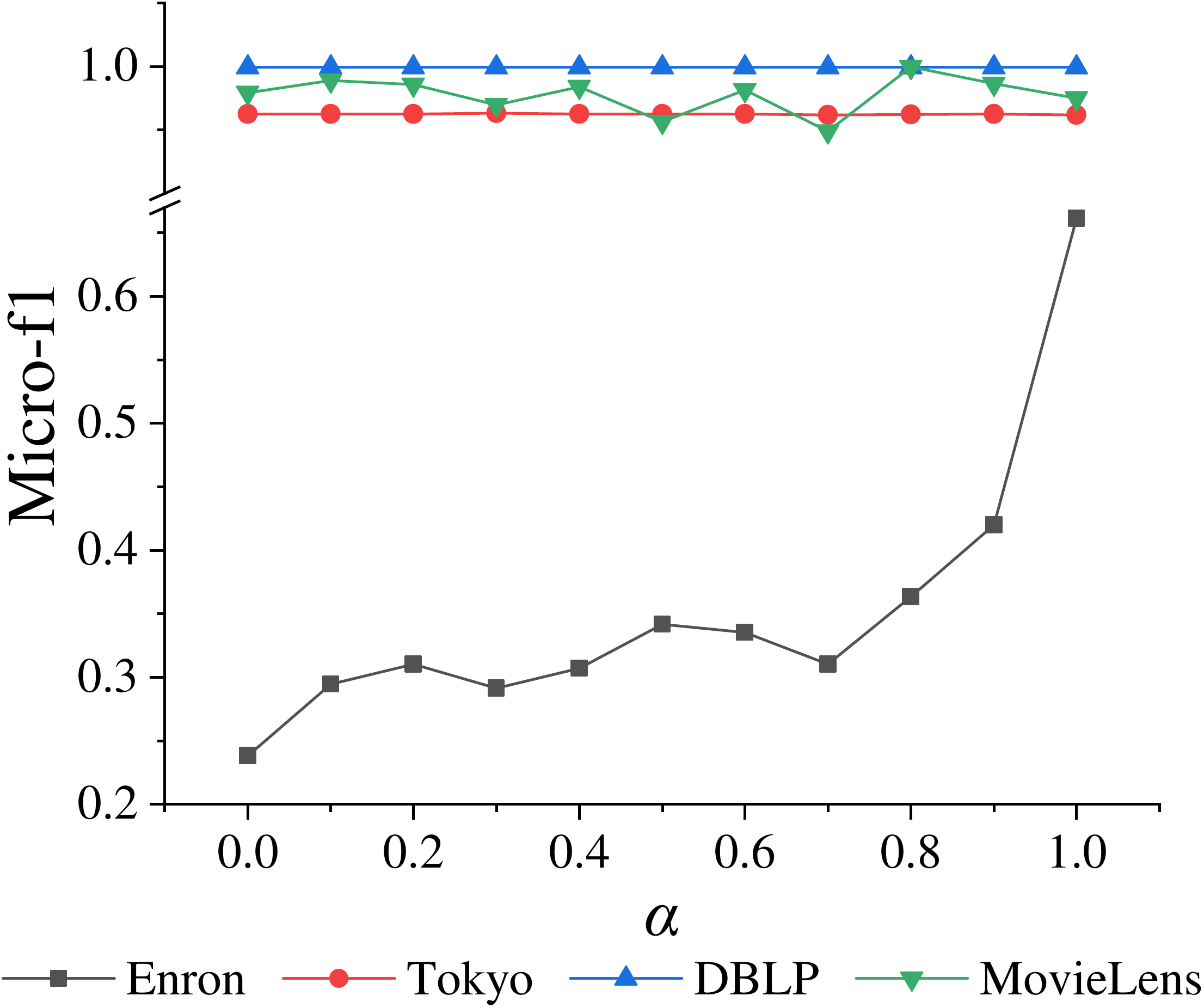}
        \end{minipage}
    }\\
    \subfigure[]{
        \begin{minipage}[t]{0.48\textwidth}
            \centering
            \includegraphics[width=0.95\textwidth,height=0.7\textwidth]{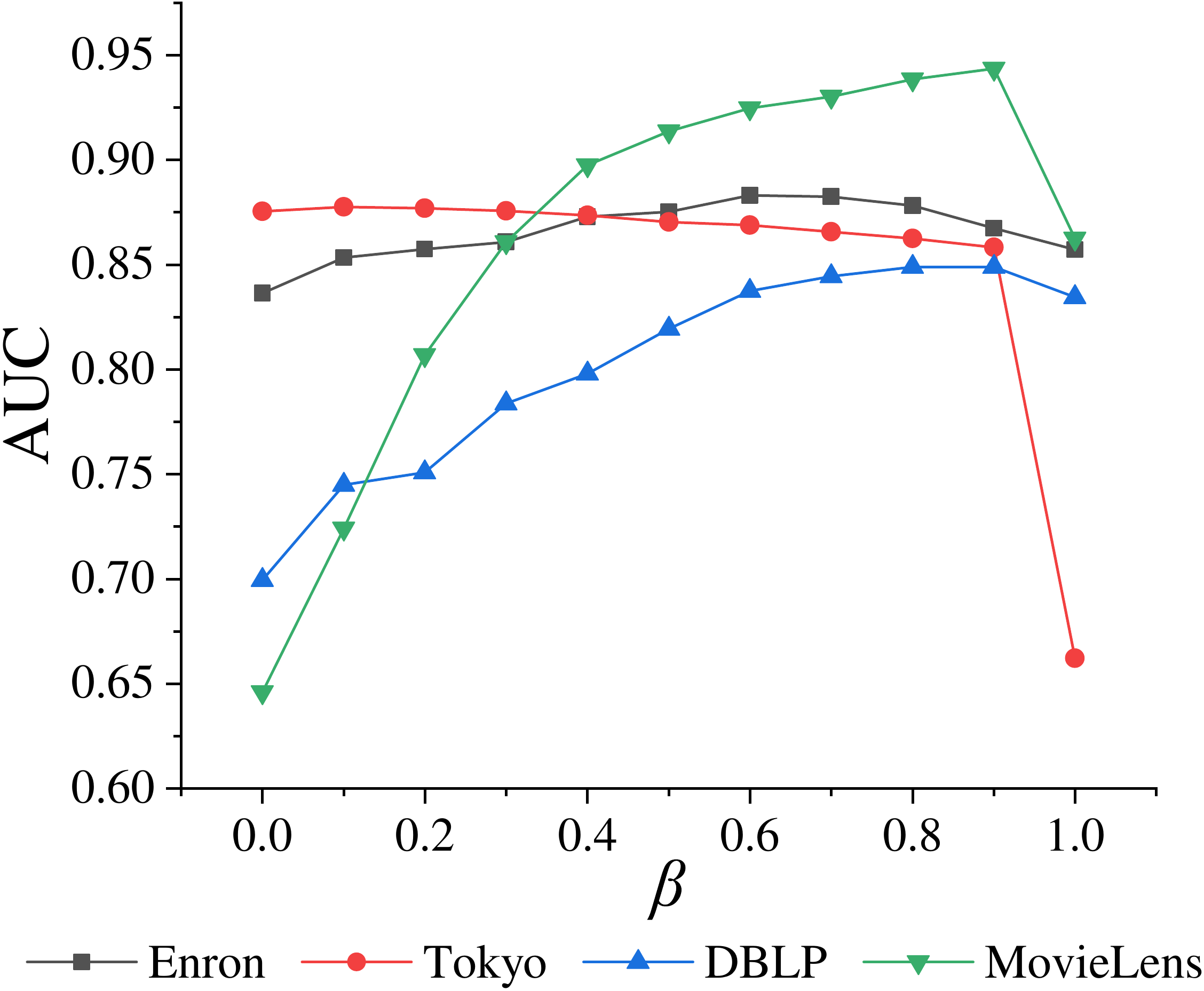}
        \end{minipage}
    }
    \subfigure[]{
        \begin{minipage}[t]{0.48\textwidth}
            \centering
            \includegraphics[width=0.95\textwidth,height=0.7\textwidth]{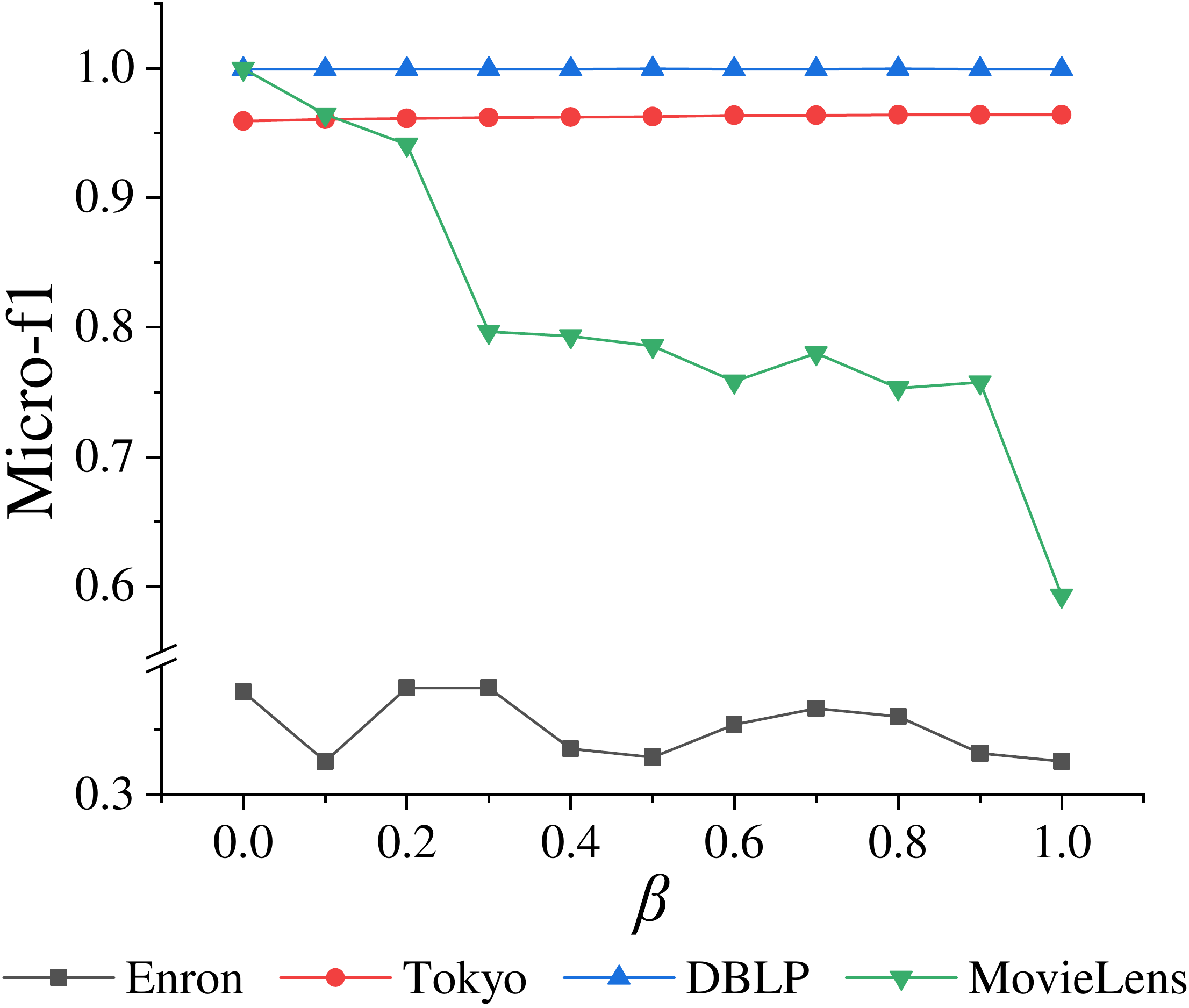}
        \end{minipage}
    }
    \caption{The influence of parameters $\alpha$ and $\beta$ for two experimental tasks in four datasets: (a) $\alpha$ for temporal link prediction, (b) $\alpha$ for node classification, (c) $\beta$ for temporal link prediction, (d) $\beta$ for node classification. The X-axis represents values of $\alpha$ and $\beta$, while the Y-axis represents AUC for temporal link prediction task and micro-f1 for node classification task. Best viewed in color.}
    \label{fig::expr:parameter}
\end{figure}

We also analyze the influence of parameters $\alpha$ and $\beta$ for two experimental tasks in four datasets. As Figure~\ref{fig::expr:parameter} shows, no matter for temporal link prediction or node classification, datasets with multiple types of nodes (e.g. Enron) are more affected by $\alpha$ while those with fine temporal granularity (e.g. MovieLens) are more affected by $\beta$, which conforms to the control of random walk strategy. According to our experience, in temporal link prediction, both $\alpha$ and $\beta$ are expected to be a larger value except 1 because this may lead to overly constraints and stop the random walk process too early. While in node classification, $\alpha=1$ could keep random walks staying at the same type of nodes and strengthen the relations among them, which is especially important in multi-class classification tasks.

\subsection{Visualizations}  \label{sec::expr:visualization}

\begin{figure}[tb!]
    \centering
    \subfigure[\scriptsize{Deepwalk}]{
        \begin{minipage}[t]{0.23\textwidth}
            \centering
            \includegraphics[width=\textwidth]{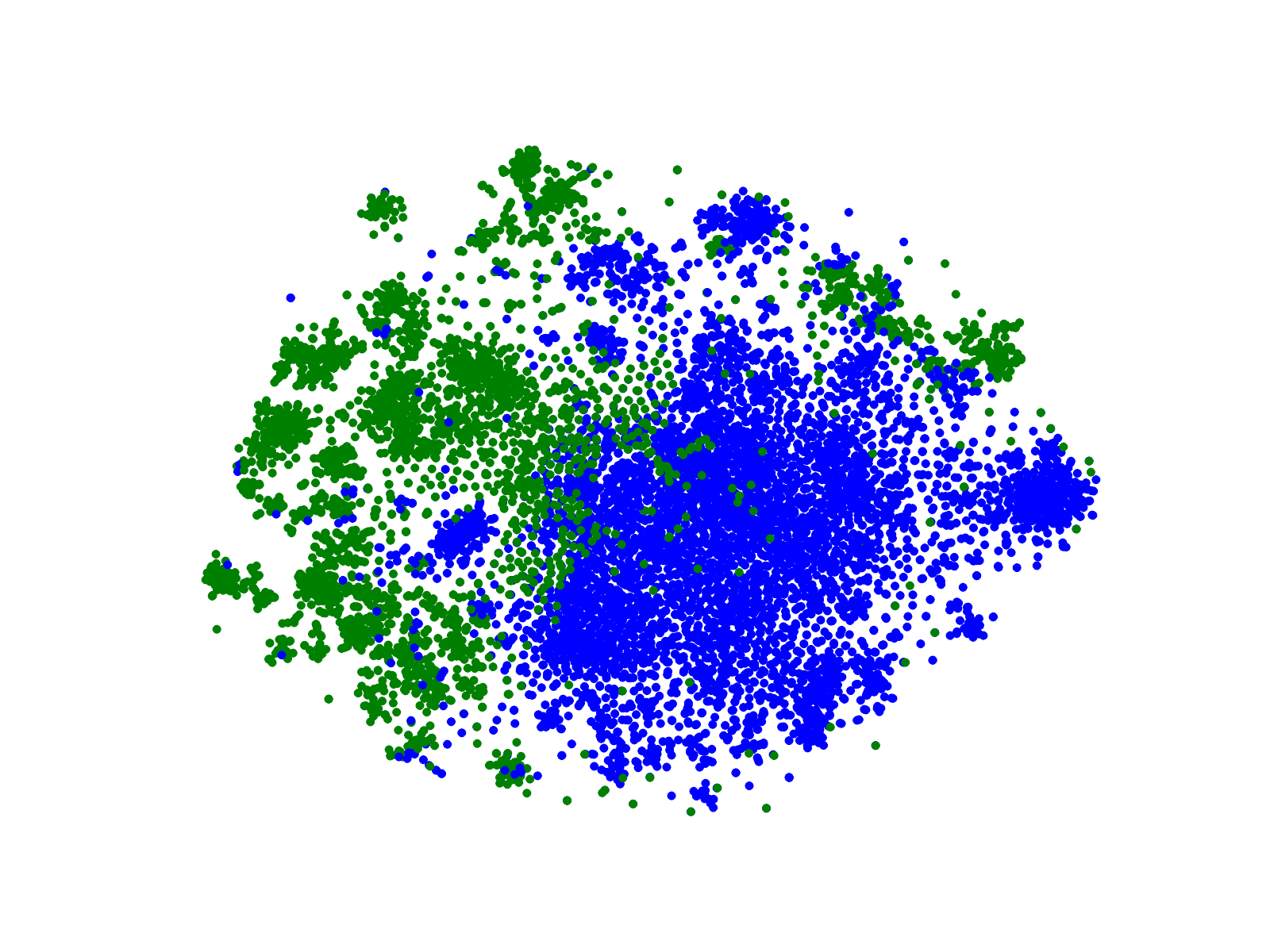}
        \end{minipage}
    }
    \subfigure[\scriptsize{CTDNE}]{
        \begin{minipage}[t]{0.23\textwidth}
            \centering
            \includegraphics[width=\textwidth]{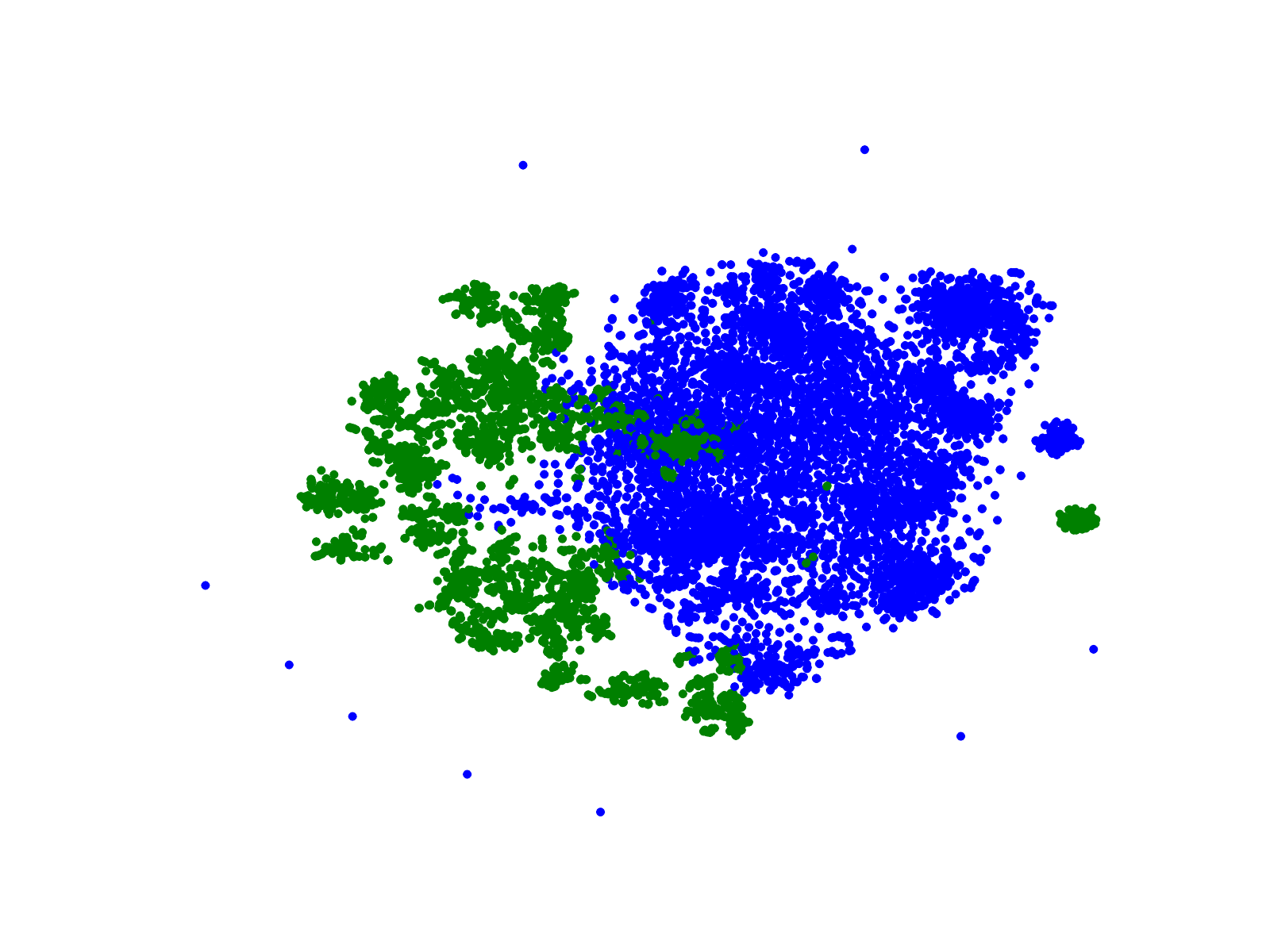}
        \end{minipage}
    }
    \subfigure[\scriptsize{ISGNS}]{
        \begin{minipage}[t]{0.23\textwidth}
            \centering
            \includegraphics[width=\textwidth]{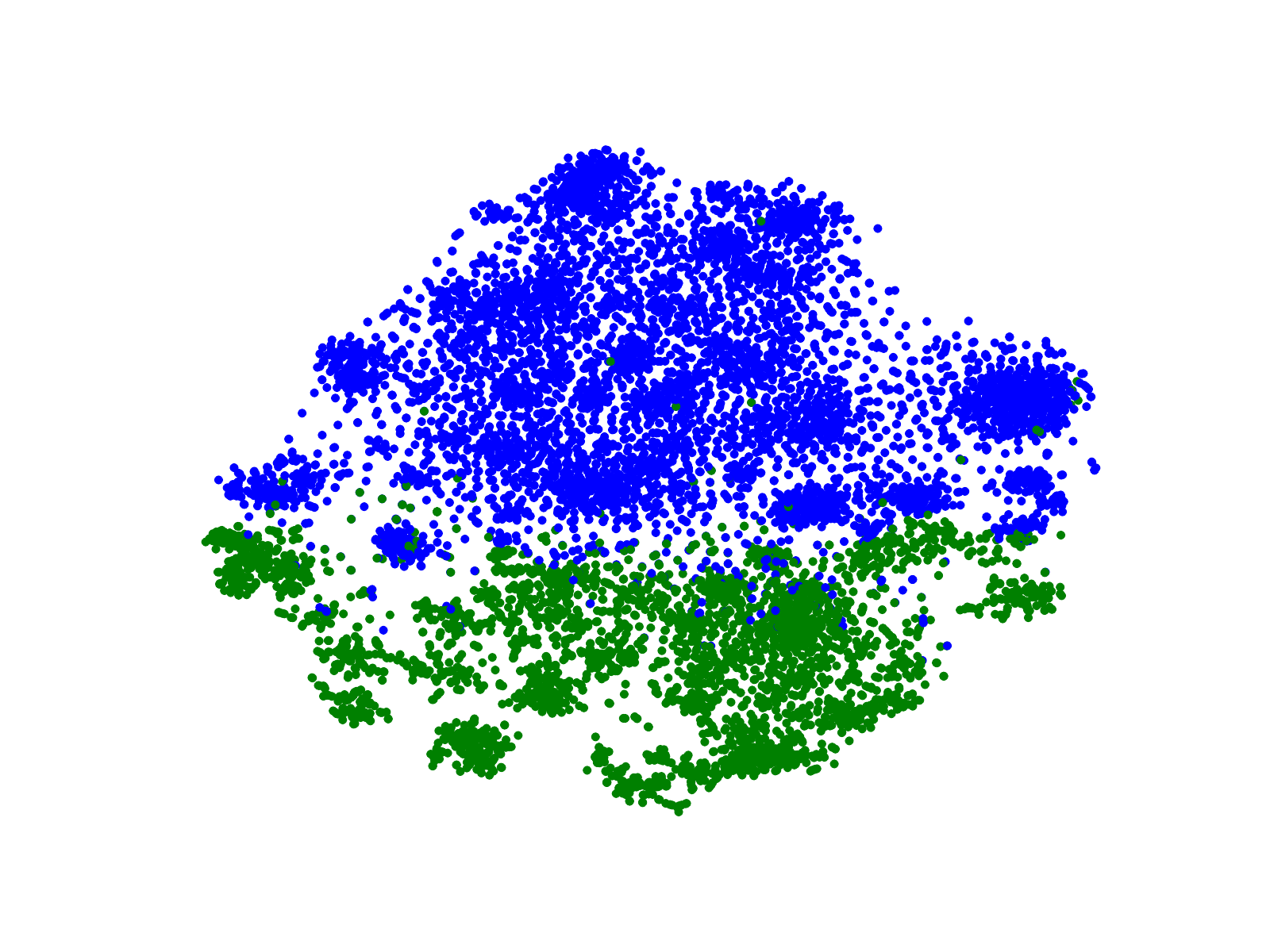}
        \end{minipage}
    }
    \subfigure[\scriptsize{Change2vec}]{
        \begin{minipage}[t]{0.23\textwidth}
            \centering
            \includegraphics[width=\textwidth]{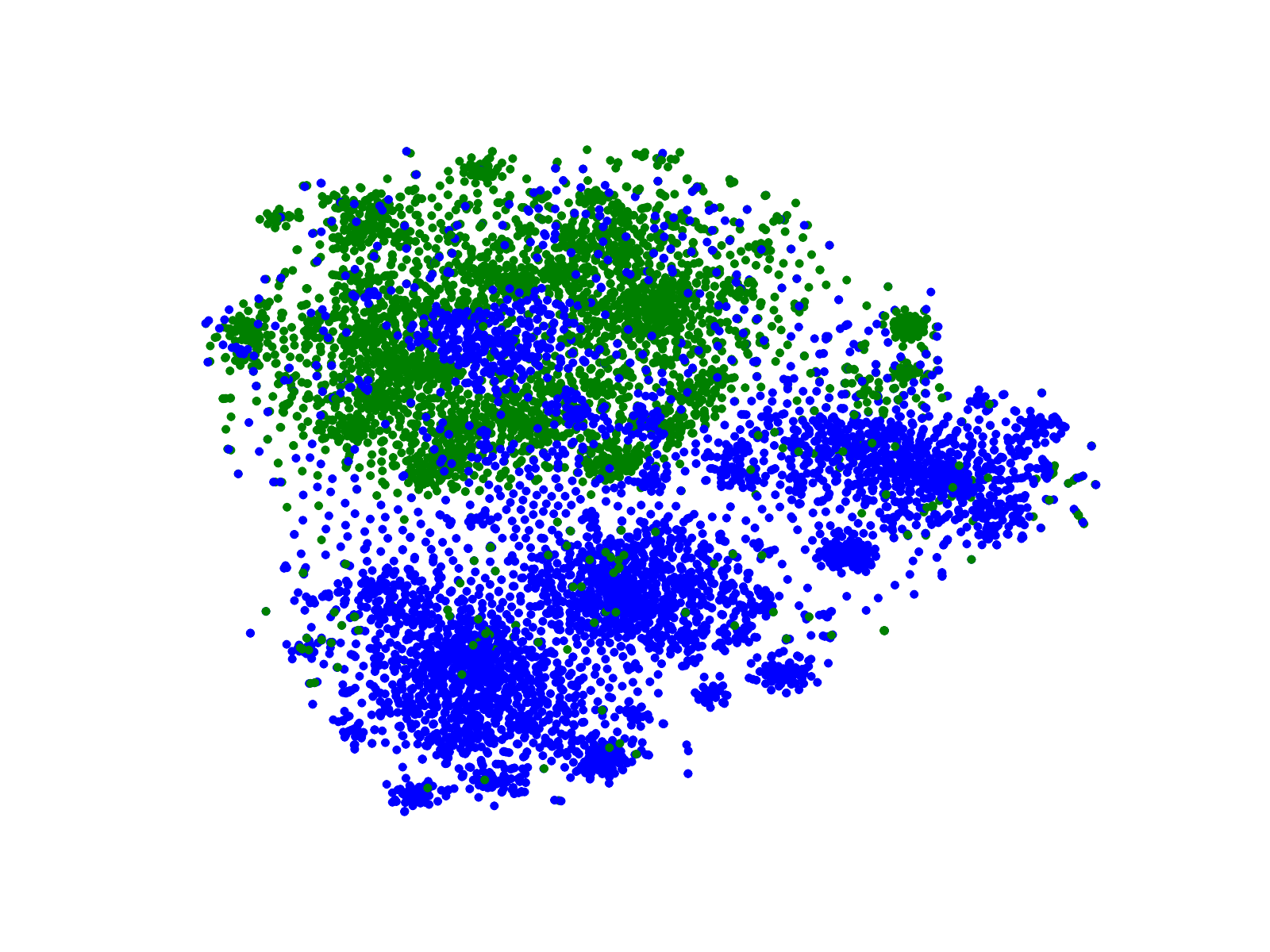}
        \end{minipage}
    }\\
    \subfigure[\scriptsize{DHNE}]{
        \begin{minipage}[t]{0.23\textwidth}
            \centering
            \includegraphics[width=\textwidth]{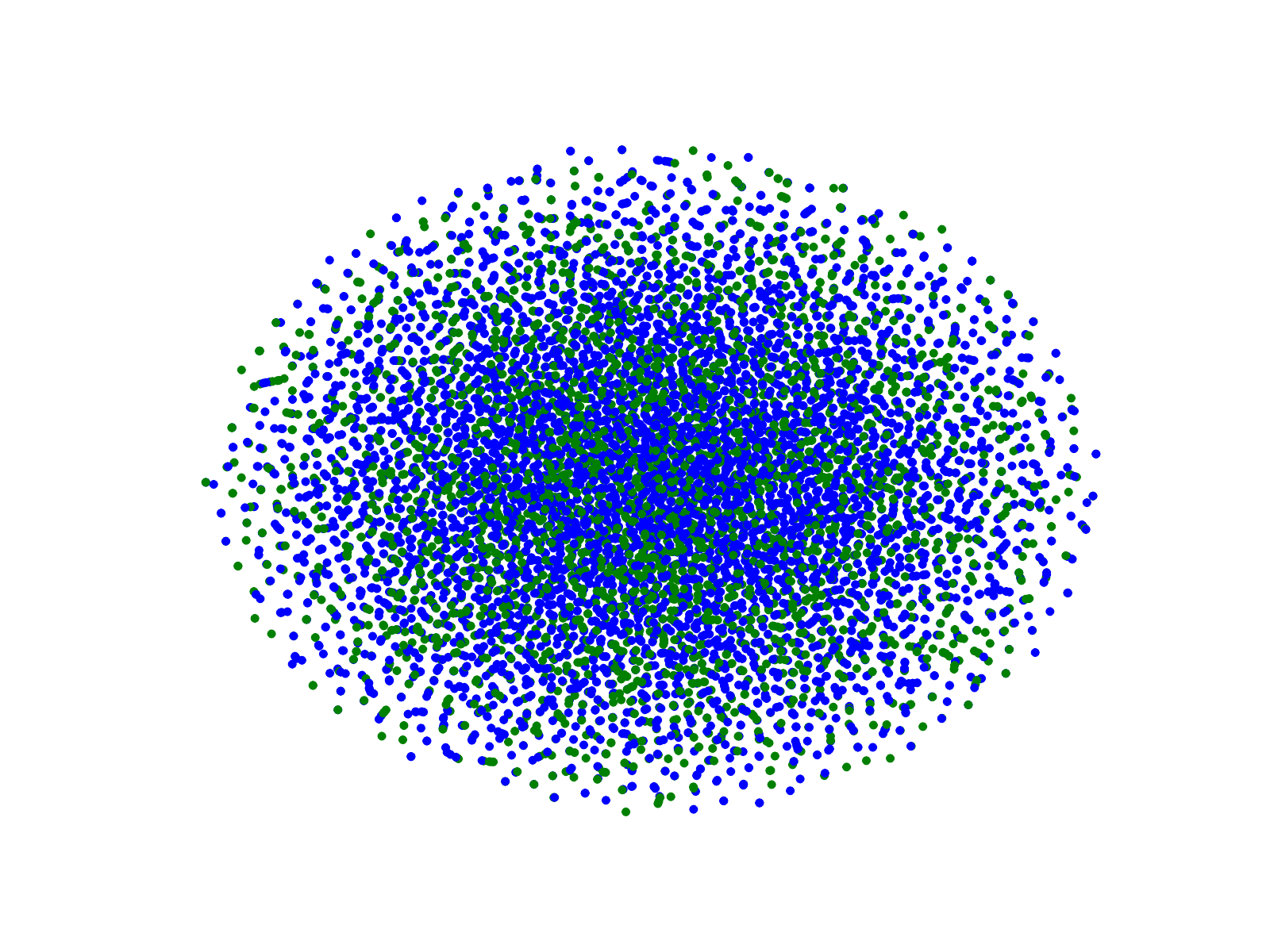}
        \end{minipage}
    }
    \subfigure[\scriptsize{HHNE}]{
        \begin{minipage}[t]{0.23\textwidth}
            \centering
            \includegraphics[width=\textwidth]{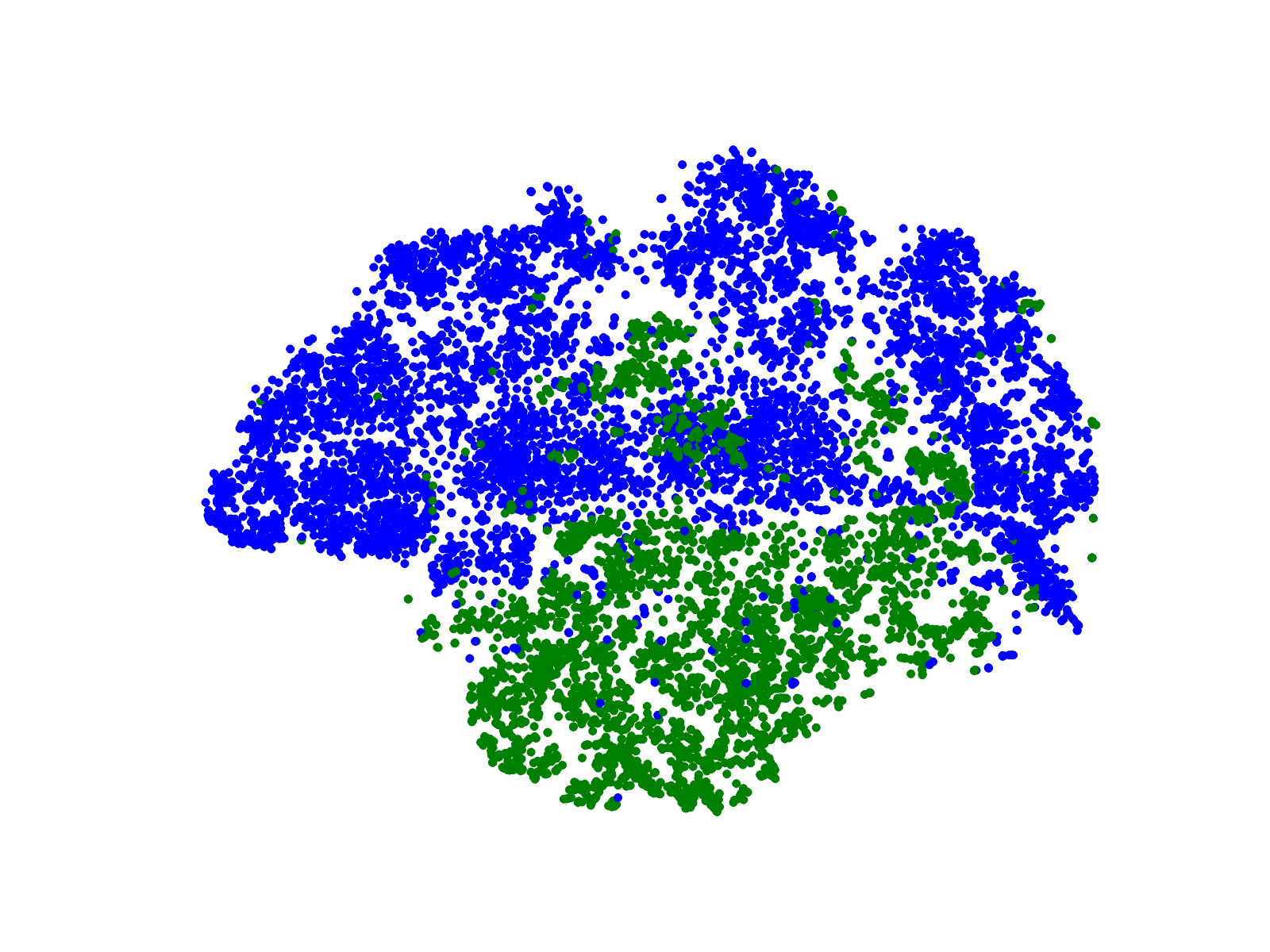}
        \end{minipage}
    }
    \subfigure[\scriptsize{GCN}]{
        \begin{minipage}[t]{0.23\textwidth}
            \centering
            \includegraphics[width=\textwidth]{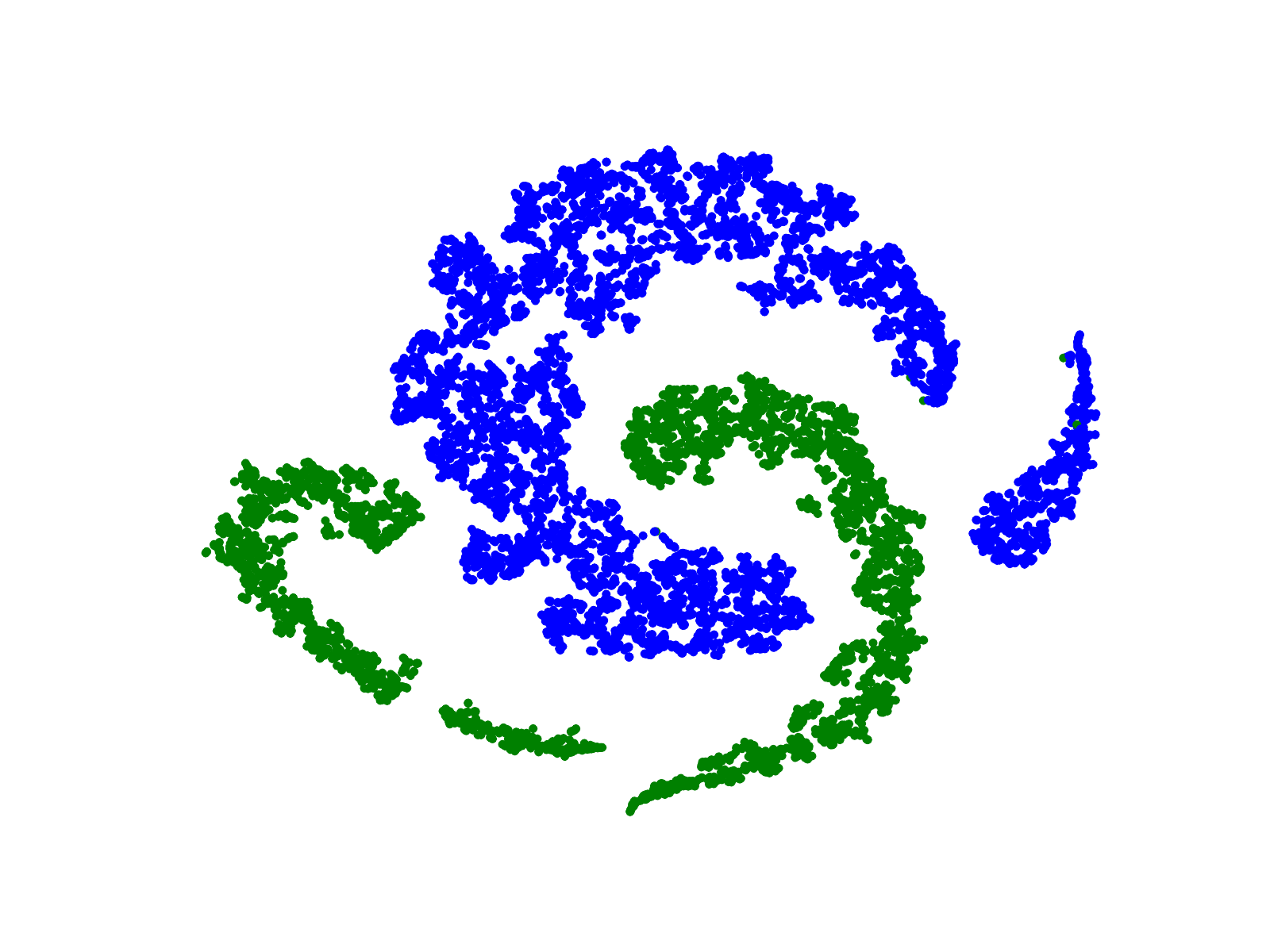}
        \end{minipage}
    }
    \subfigure[\scriptsize{GAT}]{
        \begin{minipage}[t]{0.23\textwidth}
            \centering
            \includegraphics[width=\textwidth]{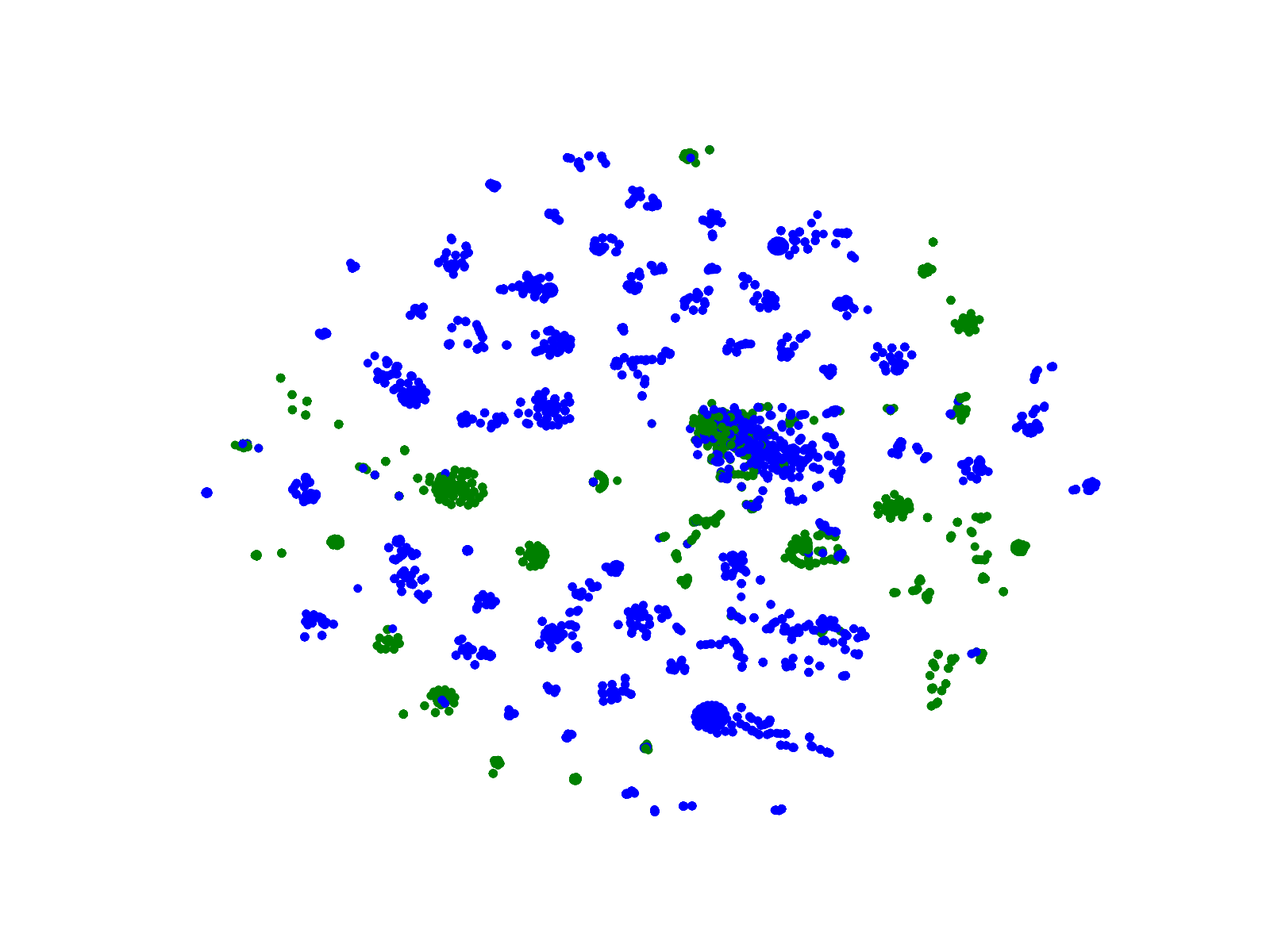}
        \end{minipage}
    }\\
    \subfigure[\scriptsize{DySAT}]{
        \begin{minipage}[t]{0.23\textwidth}
            \centering
            \includegraphics[width=\textwidth]{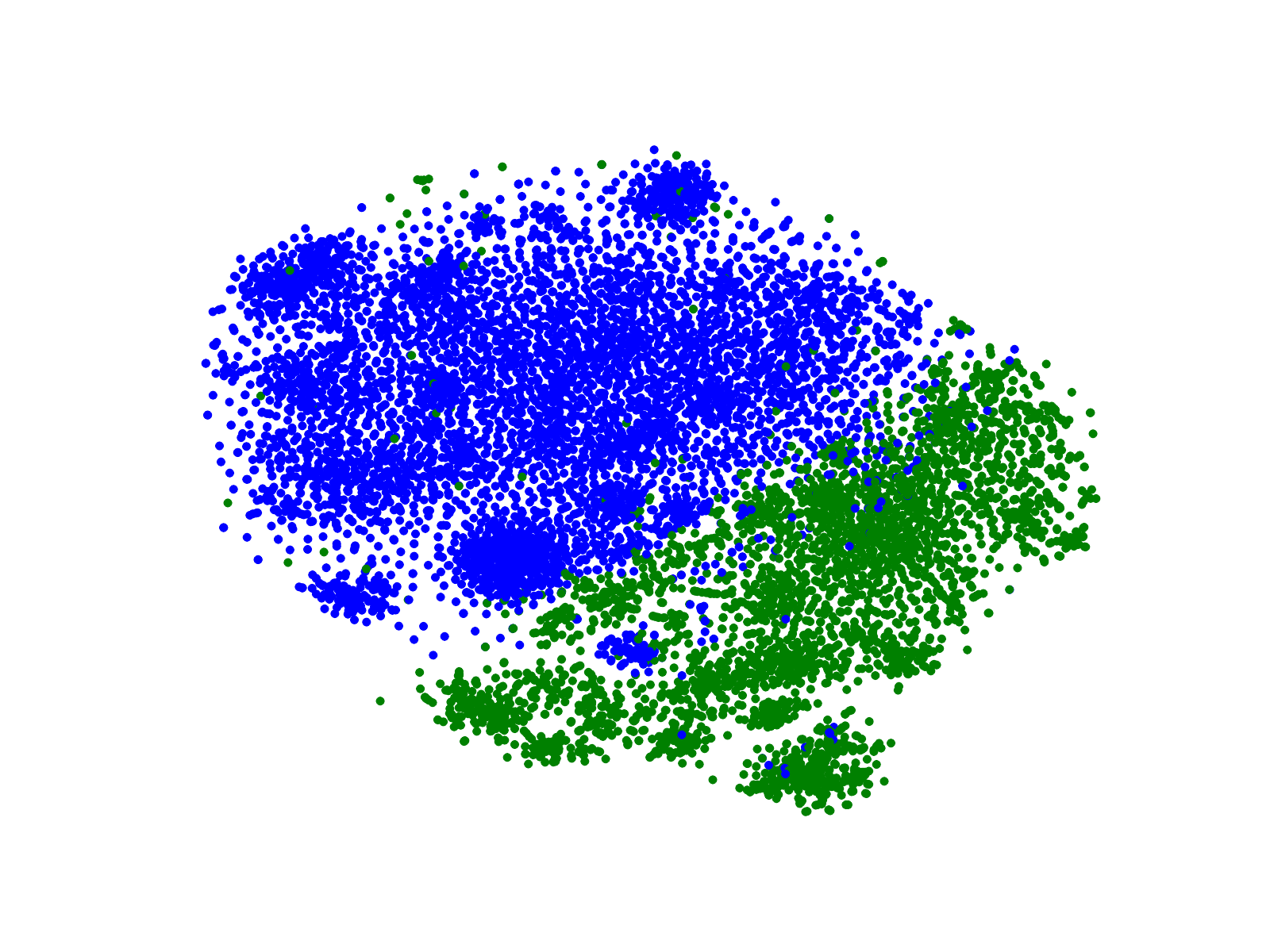}
        \end{minipage}
    }
    \subfigure[\scriptsize{LIME}]{
        \begin{minipage}[t]{0.23\textwidth}
            \centering
            \includegraphics[width=\textwidth]{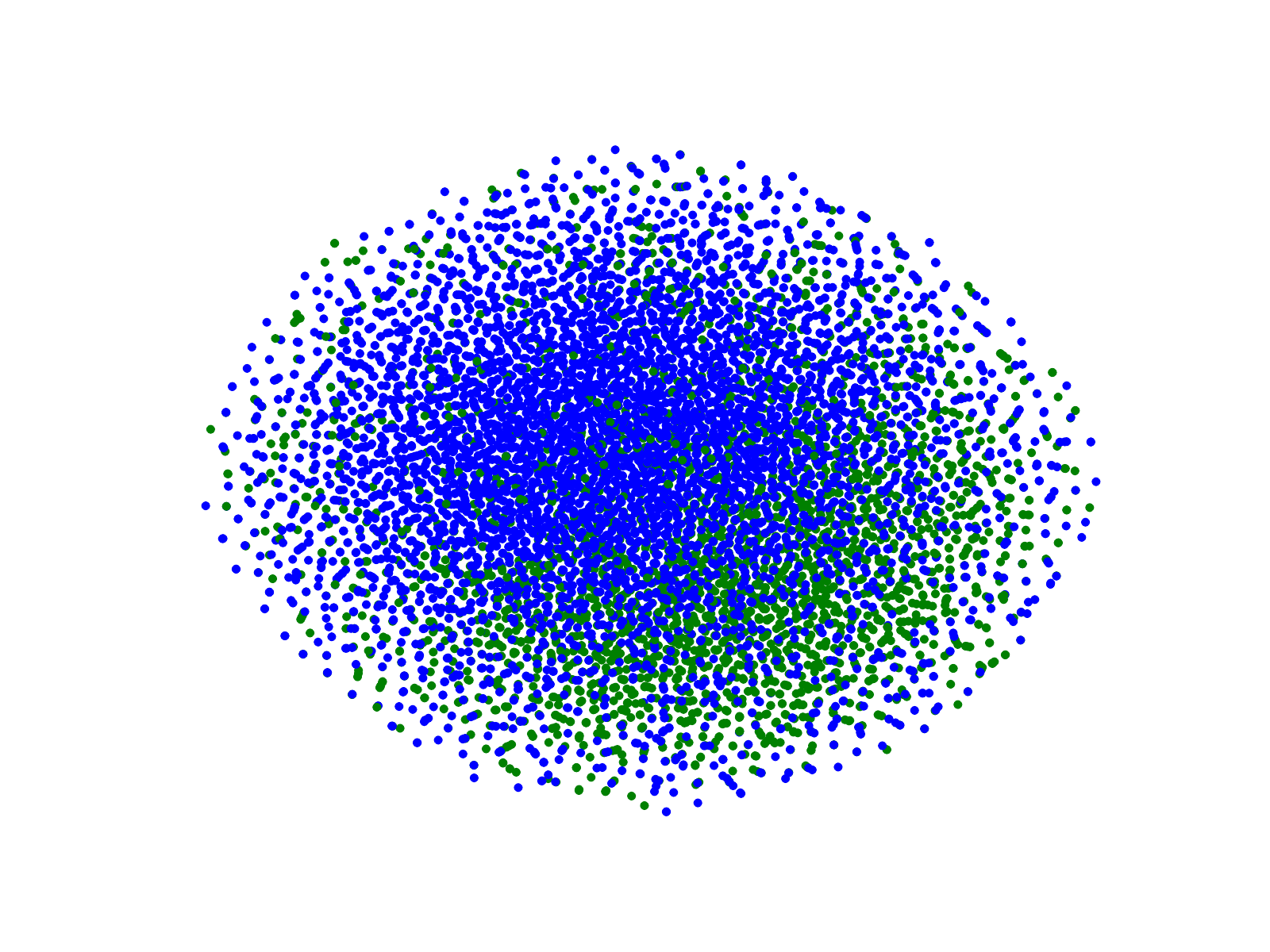}
        \end{minipage}
    }
    \subfigure[\scriptsize{TGAT}]{
        \begin{minipage}[t]{0.23\textwidth}
            \centering
            \includegraphics[width=\textwidth]{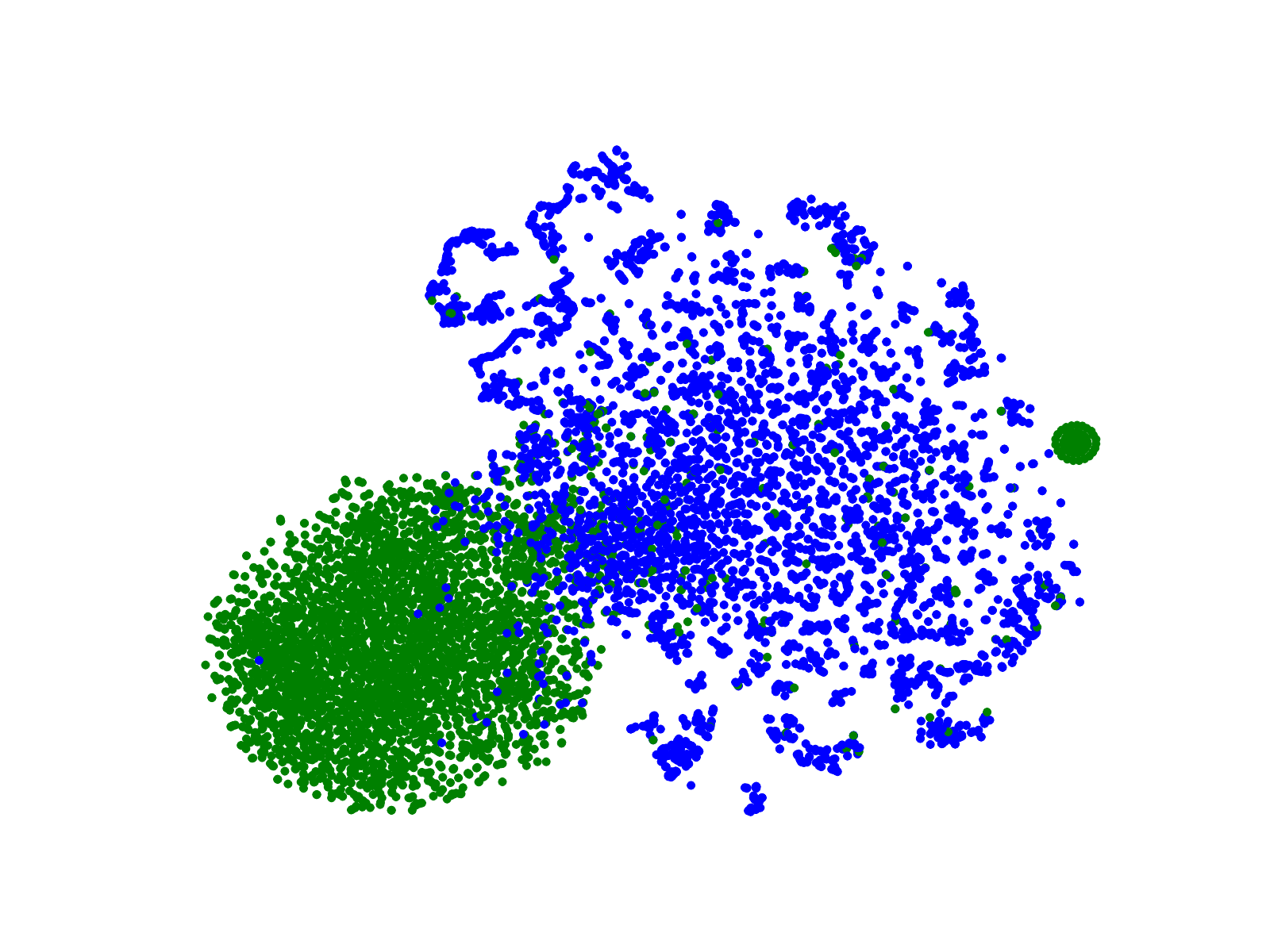}
        \end{minipage}
    }
    \subfigure[\scriptsize{\hhtne}]{
        \begin{minipage}[t]{0.23\textwidth}
            \centering
            \includegraphics[width=\textwidth]{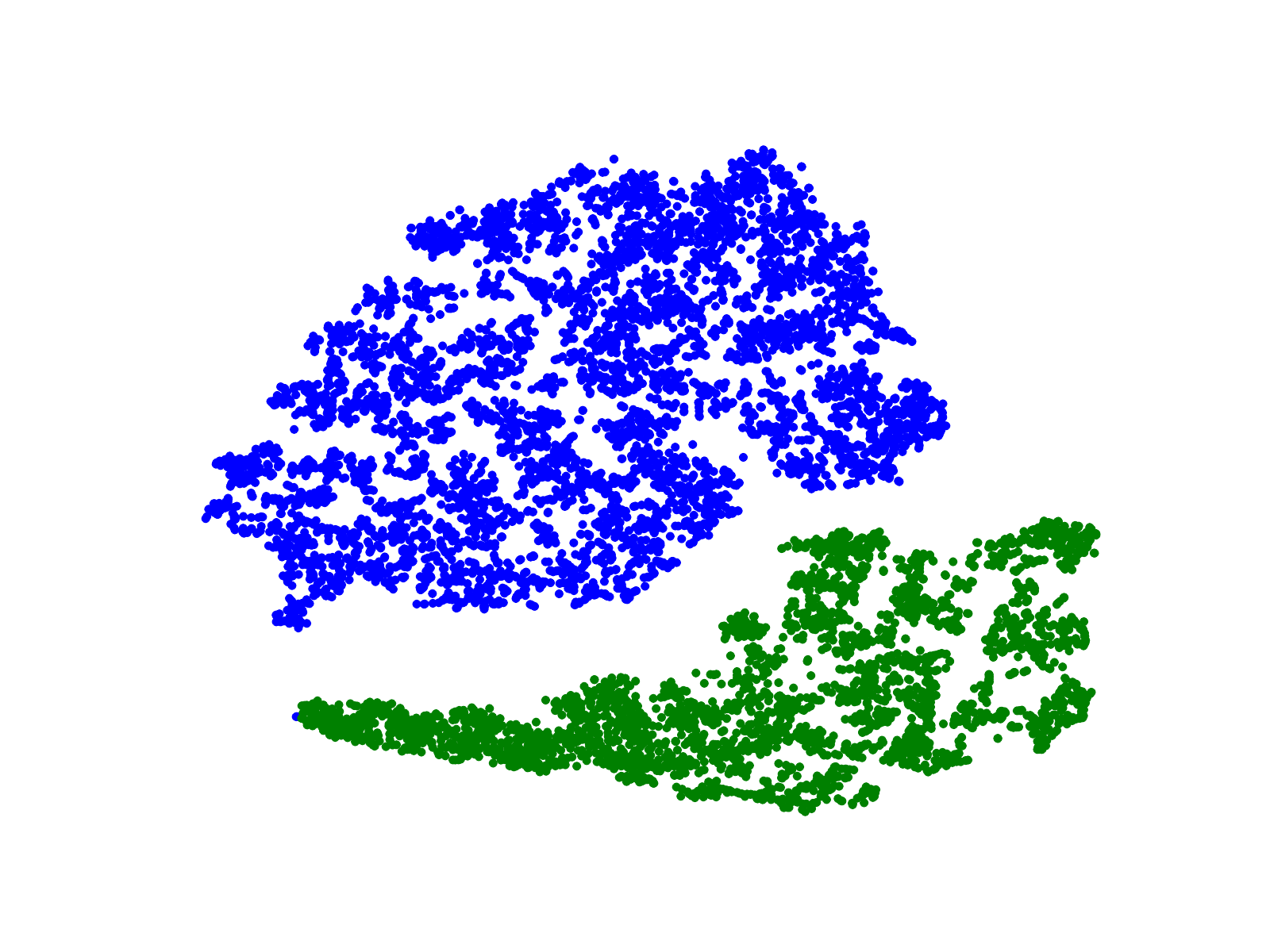}
        \end{minipage}
    }
    \caption{The visualizations of all baselines and \hhtne ~for node classification task in MovieLens. Different colors refer different types of nodes. Best viewed in color.}
    \label{fig::expr:visualization}
\end{figure}

We further study the visualizations of all nodes. Figure~\ref{fig::expr:visualization} shows the embedding layouts, which are projected to 2 dimensions from 128 by t-SNE, on dataset MovieLens for eleven baselines and our proposed \hhtne~. Different colors refer different types of nodes. Obviously our method has the best capacity of discriminating different node types compared with other models.

\section{Conclusion}  \label{sec::conclu}
In this paper, we propose a hyperbolic temporal HIN embedding model \hhtne ~based on the power-law distribution of real-world networks. \hhtne ~first leverages a temporally and heterogeneously double-constrained random walk strategy and then embeds nodes in achieved random walks into a hyperbolic space. Extensive experimental results prove the superior performance of our model.

\paragraph*{Supplemental Material Statement:} 
Section~\ref{sec::expr:setup} has provided the detailed experimental hardware and software environments, parameter settings and provenances of all datasets. Source code is available from \href{https://github.com/TaiLvYuanLiang/H2TNE}{https://github.com/TaiLvYuanLiang/H2TNE}.

\paragraph*{Acknowledgements.}
This work is supported by the Chinese Scientific and Technical Innovation Project 2030 (2018AAA0102100), National Natural Science Foundation of China (U1936206, U1903128,62172237) and the Fundamental Research Funds for the Central Universities(No. 63223046).

\newpage
\bibliographystyle{splncs04}
\bibliography{biblist}

\begin{thebibliography}{10}
\providecommand{\url}[1]{\texttt{#1}}
\providecommand{\urlprefix}{URL }
\providecommand{\doi}[1]{https://doi.org/#1}

\bibitem{bianNetworkEmbeddingChange2019a}
Bian, R., Koh, Y.S., Dobbie, G., Divoli, A.: Network embedding and change
  modeling in dynamic heterogeneous networks. In: Proceedings of the 42nd
  International ACM SIGIR Conference on Research and Development in Information
  Retrieval. pp. 861--864 (2019)

\bibitem{bronsteinGeometricDeepLearning2017}
Bronstein, M.M., Bruna, J., LeCun, Y., Szlam, A., Vandergheynst, P.: Geometric
  deep learning: going beyond euclidean data. IEEE Signal Processing Magazine
  \textbf{34}(4),  18--42 (2017)

\bibitem{chamiHyperbolicGraphConvolutional}
Chami, I., Ying, Z., R{\'e}, C., Leskovec, J.: Hyperbolic graph convolutional
  neural networks. Advances in neural information processing systems
  \textbf{32} (2019)

\bibitem{chenGraphRepresentationLearning2020}
Chen, F., Wang, Y.C., Wang, B., Kuo, C.C.J.: Graph representation learning: a
  survey. APSIPA Transactions on Signal and Information Processing  \textbf{9}
  (2020)

\bibitem{chenTutorialNetworkEmbeddings2018}
Chen, H., Perozzi, B., Al-Rfou, R., Skiena, S.: A tutorial on network
  embeddings. arXiv preprint arXiv:1808.02590  (2018)

\bibitem{cuiSurveyNetworkEmbedding2019}
Cui, P., Wang, X., Pei, J., Zhu, W.: A survey on network embedding. IEEE
  transactions on knowledge and data engineering  \textbf{31}(5),  833--852
  (2018)

\bibitem{dongMetapath2vecScalableRepresentation2017}
Dong, Y., Chawla, N.V., Swami, A.: metapath2vec: Scalable representation
  learning for heterogeneous networks. In: Proceedings of the 23rd ACM SIGKDD
  international conference on knowledge discovery and data mining. pp. 135--144
  (2017)

\bibitem{fuMAGNNMetapathAggregated2020}
Fu, X., Zhang, J., Meng, Z., King, I.: Magnn: Metapath aggregated graph neural
  network for heterogeneous graph embedding. In: Proceedings of The Web
  Conference 2020. pp. 2331--2341 (2020)

\bibitem{groverNode2vecScalableFeature2016}
Grover, A., Leskovec, J.: node2vec: Scalable feature learning for networks. In:
  Proceedings of the 22nd ACM SIGKDD international conference on Knowledge
  discovery and data mining. pp. 855--864 (2016)

\bibitem{jiawenguoJiYuFeiDiJianShiXuSuiJiYouZouDeDongTaiYiZhiWangLuoQianRu2021}
Guo, J., Bai, Q., Lin, Z., Song, C., Yuan, X.: Dynamic heterogeneous network
  embedding based on non-decreasing temporal random walk. Journal of Computer
  Research and Development  \textbf{58}(8), ~1624 (2021)

\bibitem{harperMovieLensDatasetsHistory2016}
Harper, F.M., Konstan, J.A.: The movielens datasets: History and context. Acm
  transactions on interactive intelligent systems (tiis)  \textbf{5}(4),  1--19
  (2015)

\bibitem{huangTemporalHeterogeneousInformation2021}
Huang, H., Shi, R., Zhou, W., Wang, X., Jin, H., Fu, X.: Temporal heterogeneous
  information network embedding. In: The 30th International Joint Conference on
  Artificial Intelligence (2021)

\bibitem{husseinAreMetaPathsNecessary2018a}
Hussein, R., Yang, D., Cudr{\'e}-Mauroux, P.: Are meta-paths necessary?
  revisiting heterogeneous graph embeddings. In: Proceedings of the 27th ACM
  international conference on information and knowledge management. pp.
  437--446 (2018)

\bibitem{kipfSemiSupervisedClassificationGraph2017}
Kipf, T.N., Welling, M.: Semi-supervised classification with graph
  convolutional networks. In: International Conference on Learning
  Representations (2017)

\bibitem{krioukovHyperbolicGeometryComplex2010}
Krioukov, D., Papadopoulos, F., Kitsak, M., Vahdat, A., Bogun{\'a}, M.:
  Hyperbolic geometry of complex networks. Physical Review E  \textbf{82}(3),
  036106 (2010)

\bibitem{leeDynamicNodeEmbeddings2020}
Lee, J.B., Nguyen, G., Rossi, R.A., Ahmed, N.K., Koh, E., Kim, S.: Dynamic node
  embeddings from edge streams. IEEE Transactions on Emerging Topics in
  Computational Intelligence  \textbf{5}(6),  931--946 (2020)

\bibitem{liSequenceawareHeterogeneousGraph2021}
Li, C., Hu, L., Shi, C., Song, G., Lu, Y.: Sequence-aware heterogeneous graph
  neural collaborative filtering. In: Proceedings of the 2021 SIAM
  International Conference on Data Mining (SDM). pp. 64--72. SIAM (2021)

\bibitem{liuHyperbolicGraphNeurala}
Liu, Q., Nickel, M., Kiela, D.: Hyperbolic graph neural networks. Advances in
  Neural Information Processing Systems  \textbf{32} (2019)

\bibitem{luTemporalNetworkEmbedding2019}
Lu, Y., Wang, X., Shi, C., Yu, P.S., Ye, Y.: Temporal network embedding with
  micro-and macro-dynamics. In: Proceedings of the 28th ACM international
  conference on information and knowledge management. pp. 469--478 (2019)

\bibitem{nickelPoincareEmbeddingsLearning}
Nickel, M., Kiela, D.: Poincar{\'e} embeddings for learning hierarchical
  representations. Advances in neural information processing systems
  \textbf{30} (2017)

\bibitem{nickelLearningContinuousHierarchies}
Nickel, M., Kiela, D.: Learning continuous hierarchies in the lorentz model of
  hyperbolic geometry. In: International Conference on Machine Learning. pp.
  3779--3788. PMLR (2018)

\bibitem{pengDynamicNetworkEmbedding2020}
Peng, H., Li, J., Yan, H., Gong, Q., Wang, S., Liu, L., Wang, L., Ren, X.:
  Dynamic network embedding via incremental skip-gram with negative sampling.
  Science China Information Sciences  \textbf{63}(10),  1--19 (2020)

\bibitem{pengLIMELowCostIncremental2021}
Peng, H., Yang, R., Wang, Z., Li, J., He, L., Yu, P., Zomaya, A., Ranjan, R.:
  Lime: Low-cost incremental learning for dynamic heterogeneous information
  networks. IEEE Transactions on Computers  (2021)

\bibitem{pengHyperbolicDeepNeural2021}
Peng, W., Varanka, T., Mostafa, A., Shi, H., Zhao, G.: Hyperbolic deep neural
  networks: A survey. IEEE transactions on pattern analysis and machine
  intelligence  (2021)

\bibitem{perozziDeepWalkOnlineLearning2014}
Perozzi, B., Al-Rfou, R., Skiena, S.: Deepwalk: Online learning of social
  representations. In: Proceedings of the 20th ACM SIGKDD international
  conference on Knowledge discovery and data mining. pp. 701--710 (2014)

\bibitem{salaRepresentationTradeoffsHyperbolic}
Sala, F., De~Sa, C., Gu, A., R{\'e}, C.: Representation tradeoffs for
  hyperbolic embeddings. In: International conference on machine learning. pp.
  4460--4469. PMLR (2018)

\bibitem{sankarDySATDeepNeural2020}
Sankar, A., Wu, Y., Gou, L., Zhang, W., Yang, H.: Dysat: Deep neural
  representation learning on dynamic graphs via self-attention networks. In:
  Proceedings of the 13th International Conference on Web Search and Data
  Mining. pp. 519--527 (2020)

\bibitem{shiSurveyHeterogeneousInformation2017}
Shi, C., Li, Y., Zhang, J., Sun, Y., Philip, S.Y.: A survey of heterogeneous
  information network analysis. IEEE Transactions on Knowledge and Data
  Engineering  \textbf{29}(1),  17--37 (2016)

\bibitem{tangPTEPredictiveText2015}
Tang, J., Qu, M., Mei, Q.: Pte: Predictive text embedding through large-scale
  heterogeneous text networks. In: Proceedings of the 21th ACM SIGKDD
  international conference on knowledge discovery and data mining. pp.
  1165--1174 (2015)

\bibitem{tangLINELargescaleInformation2015}
Tang, J., Qu, M., Wang, M., Zhang, M., Yan, J., Mei, Q.: Line: Large-scale
  information network embedding. In: Proceedings of the 24th international
  conference on world wide web. pp. 1067--1077 (2015)

\bibitem{velickovicGraphAttentionNetworks2018}
Veli{\v{c}}kovi{\'c}, P., Cucurull, G., Casanova, A., Romero, A., Li{\`o}, P.,
  Bengio, Y.: Graph attention networks. In: International Conference on
  Learning Representations (2018)

\bibitem{wangHeterogeneousGraphAttention2019}
Wang, X., Ji, H., Shi, C., Wang, B., Ye, Y., Cui, P., Yu, P.S.: Heterogeneous
  graph attention network. In: The world wide web conference. pp. 2022--2032
  (2019)

\bibitem{wangHyperbolicHeterogeneousInformation2019}
Wang, X., Zhang, Y., Shi, C.: Hyperbolic heterogeneous information network
  embedding. In: Proceedings of the AAAI conference on artificial intelligence.
  vol.~33, pp. 5337--5344 (2019)

\bibitem{xuInductiveRepresentationLearning2020}
Xu, D., Ruan, C., K{\"{o}}rpeoglu, E., Kumar, S., Achan, K.: Inductive
  representation learning on temporal graphs. In: International Conference on
  Learning Representations (2020)

\bibitem{dingqiyangModelingUserActivity2015}
Yang, D., Zhang, D., Zheng, V.W., Yu, Z.: Modeling user activity preference by
  leveraging user spatial temporal characteristics in lbsns. IEEE Transactions
  on Systems, Man, and Cybernetics: Systems  \textbf{45}(1),  129--142 (2014)

\bibitem{yangDiscretetimeTemporalNetwork2021}
Yang, M., Zhou, M., Kalander, M., Huang, Z., King, I.: Discrete-time temporal
  network embedding via implicit hierarchical learning in hyperbolic space. In:
  Proceedings of the 27th ACM SIGKDD Conference on Knowledge Discovery \& Data
  Mining. pp. 1975--1985 (2021)

\bibitem{yinDHNENetworkRepresentation2019}
Yin, Y., Ji, L.X., Zhang, J.P., Pei, Y.L.: Dhne: Network representation
  learning method for dynamic heterogeneous networks. IEEE Access  \textbf{7},
  134782--134792 (2019)

\bibitem{zhangWhereAreWe2021}
Zhang, S., Chen, H., Ming, X., Cui, L., Yin, H., Xu, G.: Where are we in
  embedding spaces? In: Proceedings of the 27th ACM SIGKDD Conference on
  Knowledge Discovery \& Data Mining. pp. 2223--2231 (2021)

\bibitem{zhangHyperbolicGraphAttention2021}
Zhang, Y., Wang, X., Shi, C., Jiang, X., Ye, Y.F.: Hyperbolic graph attention
  network. IEEE Transactions on Big Data  (2021)

\bibitem{zhengEntitySetExpansion2017}
Zheng, Y., Shi, C., Cao, X., Li, X., Wu, B.: Entity set expansion with meta
  path in knowledge graph. In: Pacific-Asia conference on knowledge discovery
  and data mining. pp. 317--329. Springer (2017)

\bibitem{zhouDynamicNetworkEmbedding}
Zhou, L., Yang, Y., Ren, X., Wu, F., Zhuang, Y.: Dynamic network embedding by
  modeling triadic closure process. In: Proceedings of the AAAI conference on
  artificial intelligence. vol.~32 (2018)

\end{thebibliography}

\end{document}